\documentclass[preprint,nonblindrev]{informs3} 
\UseRawInputEncoding
\OneAndAHalfSpacedXI 

\usepackage{natbib}
 \bibpunct[, ]{(}{)}{,}{a}{}{,}%
 \def\bibfont{\small}%
\usepackage{verbatim}
\usepackage{enumerate}
\usepackage{amssymb}
\usepackage{amsmath}
\usepackage{graphicx}
\usepackage{graphics}
\usepackage{subfigure}
\usepackage{setspace}
\usepackage{multirow}
\usepackage{enumerate}
\usepackage{BOONDOX-cal}
\usepackage{algorithm,algorithmic,clrscode3e,longtable}
\usepackage{hyperref}
\usepackage{url}
\usepackage{booktabs}

\makeatletter

\newcommand{\Rmnum}[1]{\expandafter\@slowromancap\romannumeral #1@}
\makeatother

\TheoremsNumberedThrough     

\EquationsNumberedThrough    

\MANUSCRIPTNO{} 
\begin{document}




\TITLE{The Electric Vehicle Routing Problem with Nonlinear Charging Functions}
\ARTICLEAUTHORS{%

\AUTHOR{Yijing Liang}
\AFF{School of Management and Engineering,Nanjing University,Nanjing 210093, P.R. China,
\EMAIL{liangyj@smail.nju.edu.cn} \URL{}}
\AUTHOR{Said Dabia} \AFF{Department of Operations Analytics, Vrije Universiteit Amsterdam, 1081 HV Amsterdam, Netherlands, \EMAIL{s.dabia@vu.nl}}
\AUTHOR{Zhixing Luo}
\AFF{School of Management and Engineering,Nanjing University,Nanjing 210093, P.R. China,
\EMAIL{luozx.hkphd@gmail.com} \URL{}}
} 

%

\ABSTRACT{
This paper outlines an exact and a heuristic algorithm for the electric vehicle routing problem with a nonlinear charging function (E-VRP-NL) introduced by \citet{montoya2017electric}. The E-VRP-NL captures several realistic features of electric vehicles including the battery limited driving range and nonlinear charging process at the charging stations. We formulate this problem as a set-partitioning and solve it using a column generation based algorithm. The resulting pricing problem of the column generation is a complicated problem as, next to the usual operational constraints e.g. time windows and vehicle capacity, electric vehicle related features are also considered. In particular, the nonlinear nature of the battery charging process requires the incorporation of a set of sophisticated recursive functions in the pricing algorithm. We show how these recursive functions allow for the simultaneous evaluation of the routing and charging decisions. Moreover, we illustrate how they can efficiently be embedded in the pricing algorithm. The column generation algorithm is integrated in a branch and bound algorithm and cutting planes are added resulting in a branch-and-price-and-cut algorithm for the E-VRP-NL. Next to the exact algorithm, we  also develop a tabu search based heuristic to solve the problem quickly. To prove the efficiency of the proposed algorithms, their performance is tested on benchmark instances from the literature. Our exact algorithm can optimally solve instances with up to 40 customers, including several instances previously unsolved to optimality. The tabu search heuristic proves to be superior to state-of-the-art heuristics in the literature both on solution quality and computation times.}

\maketitle
\section{Introduction}
Worldwide concerns regarding climate change, air pollution, sustainability and living quality are growing considerably. The transport sector is one of the biggest contributors to the negative effects on the environment in the form of greenhouse gas (GHG) emissions, noise and congestion. As such, the reduction of the greenhouse gas (GHG) emission has become a challenge that must be tackled urgently. In 2017, the transportation sector counted for 28.9\% of the total GHG emissions in the United States \citep{www.epa.gov}. In China, the Average Carbon Emission Intensity of the transportation sector is the highest, e.g 118 times larger than the financial sector \citep{www.tanpaifang.com}. To face this threatening trend, increasingly stringent environmental regulations and targets have been established in many countries. For example, the amount of carbon monoxide allowed to be emitted has been decreased by 50\% in China's Emission Standard (Stage \Rmnum{6}) \citep{kjs.mee.gov.cn}. As a result, the transport  sector has undergone tremendous changes in terms of new practices and technology. Electric Vehicles (EVs) are a potential instrument for confronting these environmental challenges and achieving the targets set by governmental regulations. However, EVs suffer from several technical performance weaknesses, especially the limited battery driving range and the long charging times. Yet, due to the rapid technological developments regarding battery driving range, the decreasing battery prices, and the long-term rising fuel costs, several leading transportation companies like Fedex, DHL, UPS and DPD have been implementing EVs for their last-mile delivery operations. This trend is confirmed by the \citet{www.eei.org}, that reported the sales of EVs increased by 79\%, 78\% and 34\% in the U.S., China and Europe, respectively, in 2018.

To boost the wider uptake of EVs, it is essential to minimize the operating cost throughout their lifetime. To further drive down their total cost of ownership, improving and developing novel planning and routing tools that take the specifics of EVs into account is indispensable. The routing of EVs concerns the determination of a set of routes with least cost that comply with both the operational constraints and EVs features. This problem is known in the literature under the name the Electric Vehicle Routing Problems (E-VRPs)(see e.g. \citet{schneider2014electric}) and fall under the more general family of Green Vehicle Routing Problems (G-VRPs) characterized by including environmental features (see e.g. \citet{erdougan2012green}). Next to E-VRPs, another stream of research considers the cost of emissions directly in the objective function, e.g. the pollution routing problems \citep{bektacs2011pollution,dabia2017exact} and the emission VRPs \citep{figliozzi2010vehicle}.

It is not straightforward to estimate the battery charging time accurately as it depends on various factors such as the State of Charge (SoC) and the charging speed. To overcome this issue, simplistic battery charging functions are assumed in the literature to allow tractable models. In particular, in most research papers dealing with E-VRPs, the charging functions are assumed to be linear or even constant in the SoC. Some studies assume batteries are recharged to their maximum capacity \citep{schneider2014electric,afroditi2014electric,preis2014energy,goeke2015routing,hiermann2016electric}. Clearly, partial charging can prevent unnecessary charging times and hence reduce routing costs as pointed out by \citet{keskin2015electric}. Consequently, other studies consider a battery partial charging policy in which the amount of energy charged is a decision variable (see e.g. \citet{keskin2015electric,bruglieri2015variable,desaulniers2016exact,montoya2015electric,
schiffer2017electric,montoya2017electric}). On the nature of the battery charging time, \citet{pelletier2017battery} illustrate that the charging process comprises two stages. During the first stage, the SoC increases linearly with the charging time. In the second stage, the speed of charging decreases and the SoC increases concavely with the charging time to avoid overcharge. Hence the charging process is governed by a nonlinear relationship between the SoC and the charging time. Dealing with such a nonlinear relationship in planning algorithms for E-VRPs is a challenging task.

Existing methods from the scientific literature cannot cope with the additional complexities stemming from nonlinear battery charging time functions. As explained by \citet{montoya2017electric}, real charging functions are governed by differential equations and incorporating  them into E-VRP models will drastically increase the computational difficulty. With this research, we aim to develop an exact method and a heuristic that capture realistic battery charging time functions, yet still remain tractable. We propose approximating the battery charging time function by means of a piecewise linear function. We must note, as pointed out by \citet{montoya2015electric}, that approximating nonlinear battery charging time functions may on one hand underestimate charging times leading to infeasible routes. On the other hand, conservative transport plans may be generated if charging times are overestimated leading to higher routing costs. Several researchers have adopted piecewise linear approximations of the battery charging time functions. However, all developed solution methods so far are heuristic based \citep{montoya2015electric,montoya2017electric,froger2017new,froger2019improved}. We are aware of one paper that directly employs nonlinear charging functions \citep{lee2020exact}.

Formally, we study the Electric Vehicle Routing Problem with Nonlinear Charging functions (E-VRP-NL). The nonlinear charging process at a charging station (CS) is approximated using a piecewise linear function. A homogeneous fleet of EVs is available to fulfil the demand of a set of geographically dispersed customers. Each EV is characterized by a limited battery driving range and storage capacity for loading cargo. For charging the batteries, a set of heterogeneous CSs with different charging functions is available. The E-VRP-NL concerns the determination of a set of routes with the least total duration where each customer is visited exactly once.  Moreover, routes are required to start and end at a depot, the total quantity served along a route does not exceed the vehicle's storage capacity and the battery driving range must be respected. Furthermore, the duration of each route must not exceed a limited allowed time. The route duration includes travel, waiting, service and charging time. We present here, to the best of our knowledge, the first exact approach for the E-VRP-NL based on a branch-and-price-and-cut (BPC) algorithm that incorporates several new features. More precisely, we developed specialized pricing procedures that efficiently handle the piecewise battery charging time functions. Moreover, new dominance procedures are introduced exploiting the problem's structure. Next, a tabu search based heuristic is also developed to solve larger instances quickly.

In summary, the scientific contributions of our paper are as follows. First, we derive exact and heuristic algorithms that efficiently deal with sophisticated recursive functions that are needed to capture a more realistic battery charging process for EVs. Another benefit of this approach is that the routing and charging decisions are performed simultaneously. Second, we propose a BPC algorithm that can solve the E-VRP-NL to optimality for the first time. A tailored label-setting algorithm involving stronger dominance rules enabling a set of labels to dominate one label is proposed for the pricing problem of the underlying column generation algorithm. Third, we develop a tabu search algorithm to solve larger instances of the problem quickly. Through extensive computational experiments, we show that our exact algorithm substantially outperforms MILP solvers and other algorithms in the literature in both the number of instances solved optimally and computational time. The tabu search heuristic  has also proven to be superior to state-of-the-art heuristics in the literature both on solution quality and computation time.

The reminder of the paper is organized as follows. In section \ref{sec:liter}, we review the literature related to the E-VRP-NL.
Section \ref{sec:problem} presents the description of the problem along with the set-partitioning formulation. Sections \ref{sec:algorithm} and \ref{sec:implementation} ouline the details of the exact algorithms, and section \ref{sec:heuristic} describes the tabu search heuristic. The computational results are presented in section \ref{sec:experiment}. Finally, section \ref{sec:conclusion} concludes the paper.

\section{Literature Review}
\label{sec:liter}
In this section, we review the literature related to the E-VRP-NL and other close problems in the family of the Green Vehicle Routing Problems (G-VRP). In particular, we review literature on routing alternative fuel-powered vehicles (AFVs), the Electric Vehicle Routing Problem (E-VRP),  the Electric Location Routing Problem (E-LRP) and the Two-Echelon Electric Vehicle Routing Problem (2E-E-VRP).

Early work on the routing of EVs lies in the field of Green Vehicle Routing Problem (G-VRP) first introduced by \citet{erdougan2012green}. The family of G-VRPs takes sustainability measures into consideration either by including sustainability related costs in the objective function \citep{bektacs2011pollution,dabia2017exact} or by adopting environmentally friendly vehicles powered by alternative energy resources such as biodiesel, electricity and hydrogen \citep{erdougan2012green,schneider2014electric}. Unlike \citet{erdougan2012green} where the refuelling time of AFVs is set to be fixed and the fully refuelling policy is employed, other studies allow partial recharges, and the charging amount is regarded as a decision variable. In \citet{felipe2014heuristic}, the G-VRP with partial recharges is studied where, at each CS, multiple charging technologies are available and the objective is to determine the charging amount and charging technology for each EV in order to minimize recharging cost. A heuristic integrating a greedy construction and a local search procedure into a simulated annealing framework is developed to solve the problem.\\
Another stream of research considers, next to partial charging, a mixed fleet where both conventional vehicles and AFVs are used for delivery. In \citet{sassi2014vehicle}, next to the mixed fleet, the charging cost is assumed to be time-dependent. The objective is to minimize the number of vehicles as well as the total cost including both travel and charging cost. \citet{macrina2019energy} further extend the problem by incorporating  energy consumption functions that depend on various factors including speed, acceleration, deceleration, load and gradients into the optimization models. They develop a hybrid large neighborhood search (HLNS) algorithm. In \citet{macrina2019green} energy consumption for conventional vehicles is calculated as the product of the distance traveled and emission factors which depend on cargo load, and is handled as constraints in the proposed model. An iterative local search is used as a solution method. To solve the G-VRP exactly, \citet{andelmin2017exact} propose an algorithm based on a multigraph where nodes represent customers, and arcs represent non-dominated paths. The results show that the algorithm can optimally solve instances with up to 110 customers. Related research dealing with fleet composition resulted in several papers (e.g. \citet{hiermann2016electric,sassi2015iterated,goeke2015routing,hiermann2019routing}).

The EVs differ substantially from the AFVs in that they require longer charging times and their driving range is susceptible to several factors such as temperature, speed, acceleration, load, and road angle \citep{pelletier201650th}. Consequently, the routing optimization of EVs requires planning algorithms that are substantially different from these used for routing AFVs. \citet{lin2016electric} study an E-VRP where the cost of travelling and energy consumption is minimized. The energy consumption is assumed to depend on the vehicle load, road angle and rolling resistance. Successful researches incorporating comprehensive energy consumption functions into routing models for EVs are performed by e.g. \citet{zhang2018electric} and \citet{ basso2019energy}. \citet{schneider2014electric} introduce the electric vehicle routing problem with time windows and recharging stations (E-VRPTW). The recharging time is assumed to be linear in the SoC and the objective is to minimize the number of employed EVs and the total traveled distance. The problem is solved by a hybrid heuristic which incorporates a variable neighborhood search algorithm in a tabu search framework. In another study, \citet{afroditi2014electric} consider the E-VRPTW where the charging time is set to be constant at each CS. Both researches require the battery of an EV is fully charged after paying a visit to a CS. \citet{keskin2016partial} relax this assumption and allow for battery partial recharges in the E-VRPTW. They solve the problem with an Adaptive Large Neighborhood Search (ALNS) algorithm. \citet{keskin2018matheuristic} further extend the problem by considering different charging rates, i.e. normal, fast and super-fast. In their ALNS algorithm, the incumbent routes are improved by a destroy and repair process followed by an enhancement procedure based on a MILP model that optimizes the charging decisions. \citet{bruglieri2015variable} propose a MILP formulation for the E-VRPTW also assuming battery partial recharging. The model aims to minimize the total cost consisting of travel, waiting and recharging time plus the number of the employed EVs. Moreover, a Variable Neighborhood Search Branching (VNSB) is developed to enable the solution of the problem in reasonable computational times. The only exact approach we are aware of for the E-VRPTW is attributed to \citet{desaulniers2016exact} who propose a BPC algorithm able to solve four variants of the problem, i.e. single vs. multiple recharges and full vs. partial recharges.

Recent research is moving towards more realistic models. In particular, a few papers consider nonlinear charging functions in the E-VRPs resulting in the E-VRP-NL. \citet{montoya2015electric,montoya2017electric} propose a node-based model for the E-VRP-NL in which the time and SoC at each node are tracked by a set of node-indexed variables. Their computational results prove the benefit of including the nonlinear charging time and partial recharges. A two-phase approach is developed: a hybrid heuristic based on an iterated local search is proposed for constructing routes, followed by a greedy procedure or a MILP to optimize battery charge-related decisions when routes are fixed. \citet{froger2019improved} further improve the route construction phase by introducing two enhanced formulations, i.e. an arc-based model and a path-based model. The nodes representing the CSs are replicated in the arc-based model to accommodate for possible multiple visits. However, the number of copies is difficult to set as excessive copies imply longer solving times, while insufficient copies may lead to an infeasible problem. To avoid replicating the CS nodes, the path-based model incorporates the concept of {\em{CS path}} which denotes the sequence of visited CSs lying between two nodes (customer or depot). A label-correcting algorithm along with dominance rules are provided to generate non-dominated CS paths. Results show that the MILP solver can solve more instances based on the path-based model. However, the second phase where the charging is decided on remains time consuming, making route evaluation procedures computationally exhaustive. \citet{lee2020exact} develop an exact method for a variant of the E-VRP-NL in which: (i) a nonlinear charging function is assumed; (ii) only a single vehicle is assigned to serve all customers; (iii) neither storage capacity nor time window constraints are considered, and (iv) the depot is also a CS and can be visited several times. A branch-and-price algorithm is developed based on an extended CS network. A route is segmented into several {\em no-charge segments} starting from and ending at the depot or CSs. A label-setting algorithm generates no-charge segments with negative reduced cost and hence avoid handling the complicated nonlinear charging functions. The master problem finds a set of segments to construct a route with minimum travel and charging time.  Additional constraints are needed in the master problem to combine no-charge segments into a feasible route. \citet{pelletier2018charge} solve the problem of deciding on charging schedules for a fleet of EVs over multiple days given routes are fixed. In the model, realistic charging functions are approximated using piecewise linear functions. In \citet{kocc2019electric}  the electric vehicle routing problem with piecewise linear charging functions is considered. Moreover, the CSs are allowed to be shared among different companies.

Other EV related features that attracted the attention of researchers are related to CSs. On the one hand, the CSs capacity is assumed to be limited and is expressed in the maximum number of EVs that can be charged simultaneously at a CS. \citet{froger2018electric} consider a limited CS capacity in the E-VRP-NL and solve the problem using a route-first assemble-second metaheuristic. An extension of this problem is studied by \citet{sweda2017adaptive} where the availability of CSs at any point in time is probabilistic. Several algorithms are introduced for finding an optimal a priori routing and recharging policy followed by solution approaches to an adaptive problem that build upon the a priori policy. On the other hand, another stream of research has focused on the Location Routing Problem that includes decisions related to the location of CSs into E-VRPs. \citet{yang2015battery} augment the problem by allowing battery swap at CSs. The objective is to minimize travel cost of the EVs and the construction cost of CSs. \citet{Hof2017Solving} propose an adaptive variable neighborhood search algorithm (AVNS) that outperforms \citet{yang2015battery} in both computation time and solution quality. \citet{arslan2016benders} use a benders decomposition approach to decide on the location of the CSs for multiple types of plug-in hybrid EVs with different ranges and that can be powered by either gasoline or electricity. \citet{schiffer2017electric} study a location routing problem with time windows and a partial recharges policy.

Finally, \citet{breunig2019electric} explore the Two-Echelon Electric Vehicle Routing Problem (2E-E-VRP) and develop an LNS-based algorithm and an exact algorithm to solve it. Conventional vehicles are employed in the first echelon while EVs are used for the second echelon. The objective is to find the set of routes with the least cost for both echelons in order to transport goods from the depot to customers through satellites. In opposite to \citet{breunig2019electric}, \citet{jie2019two} consider a 2E-E-VRP with battery swapping stations and heterogeneous EVs in both echelons. The problem is solved by a hybrid algorithm integrating a column generation algorithm and an adaptive large neighborhood search (ALNS).

\section{Problem Description and Formulation}
\label{sec:problem}
\begin{figure}[htb]
\centering
\includegraphics[scale = 0.6]{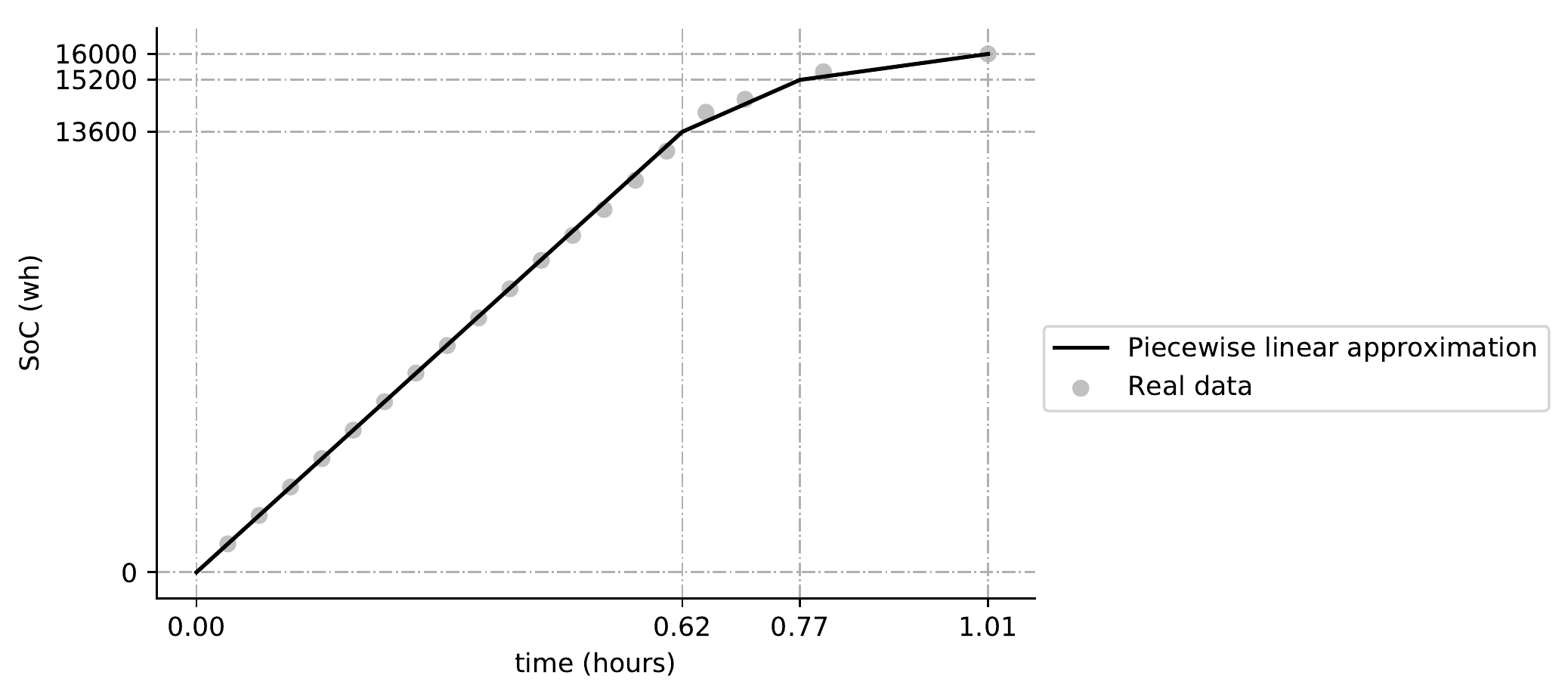}
\caption{A charging function approximated using a piecewise linear function \citep{montoya2017electric}}
\label{fig.example}
\end{figure}

Let $G = (V, A)$ be a complete digraph where $V = \{0,1,\ldots,n,n+1,\ldots,n+m\}$ is the node set and $A = \{(i,j)~|~i,j \in V,i \neq j\}$ is the arc set. Node $0$ represents the depot, the node subset $N = \{1,\ldots,n\}$ corresponds to $n$ customers, and the node subset $R = \{n+1,\ldots,n+m\}$ corresponds to $m$ recharging stations. Each customer $i \in N$ requires a service time $s_i$ and has a demand $q_i$. Furthermore,  traversing arc $(i,j)\in A$ requires a travel time $t_{i,j}$ and an amount $b_{i,j}$ of energy. The travel time matrix is supposed to satisfy the triangle inequality. We assume a limited fleet of $K$ homogenous electric vehicles is available with a limited storage capacity $Q$ and a limited battery driving range $B$. Each vehicle is required to complete its route within a time limit $T$. We denote $[e_0, l_0] = [0, T]$ the depot's time window, and $[e_i, l_i] = [t_{0,i}, T - s_i - t_{i,0}]$ the time window of customer $i \in N$. Similar to \citet{montoya2017electric}, the charging process consists of two stages. In the first stage the battery is charged at a constant rate, while in the second stage the SoC increases concavely with the charging time. This nonlinear charging process is approximated by a piecewise linear function. We assume heterogeneous CSs, where each CS $i\in R$ is characterized by a different non-decreasing piecewise linear charging function $r_i(t)$. Figure \ref{fig.example} illustrates an example of a nonlinear charging function approximated by a  piecewise function.

Our objective is to determine the set of routes with the least total duration such that: 1) the quantity moved along a route does not exceed the EV's storage capacity $Q$; 2) the battery level of the EV along a route must remain between $0$ and $B$; 3) every customer can only be served by one EV during its corresponding time window and 4) the duration of each trip must not exceed the time limit $T$. We note that a route duration comprises travel, waiting, service and charging times.

\subsection{The master problem}
To derive the set partitioning formulation for the E-VRP-NL,  we  define $\Omega$ as  the  set  of  feasible  routes. A route is feasible if it satisfies time window, storage capacity, battery driving range and duration time limit constraints. We let $c_{p}$ denote the duration (i.e. cost) of the route $p \in \Omega$ . For each route $p \in \Omega$, we let $\alpha_{i,p}$ be a constant that counts the number of times node $i$ is visited by the route $p$. Furthermore, $\theta_p$ is a binary variable that takes the value 1 if and only if the route $p$ is included in the solution, the E-VRP-NL is formulated as the following set-partitioning problem:
\begin{align}
&\min \sum_{p \in \Omega}c_p \theta_p,  \label{eq:obj} \\
\mbox{s.t.}~~&\sum_{p \in \Omega} \theta_p \leq K, \label{eq:vehicle}\\
&\sum_{p \in \Omega}\alpha_{i,p} \theta_p = 1, \quad \forall i \in N,  \label{eq:customer} \\
&\theta_p \in \{0, 1\}, \quad \forall p \in \Omega.  \label{eq:variable}
\end{align}
The  objective  function (\ref{eq:obj}) minimizes the total duration of the chosen routes. The number of employed vehicles is limited to $K$ in constraint (\ref{eq:vehicle}), and constraints (\ref{eq:customer}) guarantee that each node is visited exactly once. We use column generation to solve the linear programming (LP) relaxation of problem (\ref{eq:obj})-(\ref{eq:variable}): starting with a small subset $\Omega' = \Omega$ of variables,  we  generate  additional  variables  for  the  master  problem (the LP relaxation of (\ref{eq:obj})-(\ref{eq:variable})) by  solving a pricing problem that searches for variables with negative reduced cost. The column generation procedure terminates when no new variables with negative reduced cost are found.

\subsection{The pricing problem}
The pricing problem for the E-VRP-NL is an extension of the elementary shortest path problem with resource constraints (ESPPRC). Let $\mu_0$ and $\mu_i$ $(i \in N)$ be the dual values of constraints (\ref{eq:vehicle}) and (\ref{eq:customer}), respectively. The objective function of the pricing problem is formulated as follows:
\begin{align}
\min_{p \in \Omega} c_{p}-\sum_{i \in N}\alpha_{i,p}\mu_i - \mu_0, \label{eq:reduce}
\end{align}
The objective function (\ref{eq:reduce}) seeks the route with the least reduced cost. If the objective value of the pricing problem is nonnegative, the current solution of the master problem is optimal. We solve the pricing problem by a bounded bi-directional label-setting algorithm. For some good papers applying a (bi-directional) label-setting algorithm, we refer the reader to \cite{feillet2004exact,costa2019exact, righini2006symmetry} and \cite{boland2006accelerated}. The bounded bi-directional search split the extension process into a {\em forward extension} which extends from the (start) depot to its successors and a {\em backward extension} which extends from the (end) depot to its predecessors. The forward and backward labels are merged to construct complete routes. The  basic  operation  in  the  label-setting  algorithm is the extension of an existing label along an arc to generate a new label. These label extensions involve recursive functions that calculate the states of the newly generated labels. These recursive functions are more complicated than the ones used for the standard ESPPRC and play a key role in our algorithm. Next, we describe in more details the label-setting algorithm for both the forward and backward search including the specialized recursive functions.

\section{The Pricing Algorithm}
\label{sec:algorithm}
In this paper, we propose an exact algorithm based on column generation to solve the E-VRP-NL. In this section, we focus on the pricing algorithm which is a label-setting algorithm used to solve the pricing problem. In particular, we introduce the specialized recursive functions which are necessary to capture and handle the charging decisions. We show how these functions are embedded in the label-setting algorithm, and discuss how forward and backward labels are merged into complete routes.

\subsection{The forward recursive functions}
\label{subsec:recursive}
Let's consider a partial path $p = (i_0,\ldots,i_k)$ where $i_0 = 0$. We denote $a_{i_k}$ the earliest departure time from node $i_k$, and let $f_{i_k}(t)$ be a function representing the maximum battery level of the vehicle when leaving node $i_k$ at time $t$. We further introduce the function $r^{-1}_{i_k}(x)$, the inverse function of $r_{i_k}(t)$, representing the time required to charge an exhausted battery to level $x$. Now, we define the function $\tau_{i_k}(y) = \min\{t|f_{i_k}(t) \geq y\}$ as the earliest departure time at $i_k$ that guarantees a battery level $y$ when leaving $i_k$. In the sequel, we elaborate on the recursive calculations for $a_{i_k}$ and $f_{i_k}(t)$ where we differentiate between two cases:

\begin{enumerate}[\text{Case} 1.]
\item Node $i_k$ is a customer or the depot\\
On the one hand, we know that the arrival time at $i_k$ cannot be earlier than the lower bound of its time window $e_{i_k}$. On the other hand, the earliest arrival time at $i_k$ can be calculated as $\tau_{i_{k-1}}(b_{i_{k-1},i_k}) + t_{i_{k-1},i_k}$ where $b_{i_{k-1},i_k}$ is the minimum battery level required at $i_{k-1}$ that guarantees the vehicle reaches $i_k$ without running out of battery. Hence, the earliest departure time $a_{i_k}$ from $i_k$ is expressed as $\max\{e_{i_k}, \tau_{i_{k-1}}(b_{i_{k-1},i_k}) + t_{i_{k-1},i_k}\} + s_{i_k}$.\\
The maximum battery level $f_{i_k}(t)$ can be calculated as $f_{i_{k-1}}(t^\prime) - b_{i_{k-1},i_k}$, where $t^\prime$ is the departure time from $i_{k-1}$. The time $t^\prime$ is computed as $t - s_{i_k} - t_{i_{k-1},i_k}$, but can also be no later than $l_{i_{k-1}}+s_{{i_{k-1}}}$. Therefore, $f_{i_k}(t) = f_{i_{k-1}}(\min\{t - s_{i_k} - t_{i_{k-1},i_k}, l_{i_{k-1}}+ s_{{i_{k-1}}}\}) - b_{i_{k-1},i_k}$.

\item Node $i_k$ is a CS\\
The arrival time at $i_k$ is computed as the sum of the earliest departure time at $i_{k-1}$ and the travel time from $i_{k-1}$ to $i_k$. So we have $a_{i_k} = \tau_{i_{k-1}}(b_{i_{k-1},i_k}) + t_{i_{k-1},i_k}$.\\
The maximum battery level $f_{i_k}(t)$ is equal to the level when leaving node $i_{k-1}$ augmented by the amount charged at $i_k$. If we let the departure time at $i_{k-1}$ be $t'$, then $t'$ lies in the interval $[\tau_{i_{k-1}}(b_{i_{k-1},i_k}), t-t_{i_{k-1},i_k}]$. Thus, the charging time at $i_k$ is $t-t'-t_{i_{k-1},i_k}$. The battery level when the vehicle arrives at $i_k$ can be computed as $f_{i_{k-1}}(t') - b_{i_{k-1},i_k}$. Moreover, the maximum battery level $f_{i_k}(t)$ cannot exceed the battery capacity $B$. Consequently, $f_{i_k}(t) = \min\left\{\max_{t' \in[\tau_{i_{k-1}}(b_{i_{k-1},i_k}), t-t_{i_{k-1},i_k}]}r_{i_k}\left(r^{-1}_{i_k}\left(f_{i_{k-1}}(t') - b_{i_{k-1},i_k}\right) + t - t_{i_{k-1},i_k} - t'\right), B\right\}$.
\end{enumerate}

In the sequel, we show how $f_{i_k}(t)$ and $a_{i_k}$ can be calculated recursively. We first prove the following lemma.
\begin{lemma}
\label{lemma:1}
The function $f_{i_k}(t)$ is a non-decreasing piecewise linear function.
\end{lemma}
\textit{Proof of Lemma~\ref{lemma:1}:}
Let $t_1$, $t_2$ be two possible departure times at $i_k$ such that $t_2 \geq t_1$. We need to show that $f_{i_k}(t_2) \geq f_{i_k}(t_1)$ and do this by induction. First, we consider the first visited node $i_1$ along partial path $p$. If $i_1$ is a customer, we have $f_{i_1}(t) = f_{i_0}(\min\{t - s_{i_1} - t_{i_0,i_1}, l_{i_0} + s_{i_0}\}) - b_{i_0,i_1} = B - b_{i_0,i_1}$. The function $f_{i_1}(t)$ is constant and hence non-decreasing. If $i_1$ is a CS, then $f_{i_1}(t) = \min\left\{\max_{x \in[\tau_{i_0}(b_{i_0,i_k}), t-t_{i_0,i_k}]}r_{i_k}\left(r^{-1}_{i_k}\left(f_{i_0}(x) - b_{i_0,i_k}\right) + t - t_{i_0,i_k} - x\right), B\right\} = \min\left\{r_{i_k}\left(r^{-1}_{i_k}\left(B - b_{i_0,i_k}\right) + t - t_{i_0,i_k} - \tau_{i_0}(b_{i_0,i_k})\right), B\right\}$ which is a non-decreasing piecewise linear function because the function $r_{i_k}(t)$ is a non-decreasing piecewise linear function.\\
We now assume that $f_{i_{k-1}}(t)$ is a non-decreasing piecewise linear function, and differentiate  between two cases.\\
First, if $i_k$ is a customer, we have
\begin{align*}
\!f_{i_k}(t_2) - f_{i_k}(t_1) = f_{i_{k-1}}(\min\{t_2 - s_{i_k} - t_{i_{k-1},i_k}, l_{i_{k-1}} + s_{i_{k-1}}\})
- f_{i_{k-1}}(\min\{t_1 - s_{i_k} - t_{i_{k-1},i_k}, l_{i_{k-1}} + s_{i_{k-1}}\}) \geq 0.\!
\end{align*}
Second, if $i_k$ is a CS, we have
\begin{align*}
\!f_{i_k}(t_2) - f_{i_k}(t_1)
= \min\left\{r_{i_k}(\hat{f}_{i_k}(t_2) + t_2 - t_{i_{k-1},i_k}), B\right\} - \min\left\{r_{i_k}(\hat{f}_{i_k}(t_1) + t_1 - t_{i_{k-1},i_k}), B\right\},\!
\end{align*}
where $\hat{f}_{i_k}(t) = \max_{x \in[\tau_{i_{k-1}}(b_{i_{k-1},i_k}), t-t_{i_{k-1},i_k}]}\left\{r^{-1}_{i_k}\left(f_{i_{k-1}}(x) - b_{i_{k-1},i_k}\right) - x\right\}$.\\
Since $r_{i_k}(t)$ is a non-decreasing piecewise linear function, we only need to verify that $\hat{f}_{i_k}(t_2) + t_2 - t_{i_{k-1},i_k}-(\hat{f}_{i_k}(t_1) + t_1 - t_{i_{k-1},i_k})$ is non-negative. We have
\begin{small}
\begin{equation*}
\begin{split}
\hat{f}_{i_k}(t_2) + t_2 - t_{i_{k-1},i_k}-(\hat{f}_{i_k}(t_1) + t_1 - t_{i_{k-1},i_k}) &= \max_{x \in[\tau_{i_{k-1}}(b_{i_{k-1},i_k}), t_2-t_{i_{k-1},i_k}]}\left\{r^{-1}_{i_k}\left(f_{i_{k-1}}(x)- b_{i_{k-1},i_k}\right)- x\right\}\\
&- \max_{x \in[\tau_{i_{k-1}}(b_{i_{k-1},i_k}), t_1-t_{i_{k-1},i_k}]}\left\{r^{-1}_{i_k}\left(f_{i_{k-1}}(x) - b_{i_{k-1},i_k}\right) - x\right\} + t_2 - t_1 \\
&\geq 0,
\end{split}
\end{equation*}
\end{small}
as desired $\square$

Now, $a_{i_k}$ and $f_{i_k}(t)$ can be computed according to the following recursive operations:
\begin{small}
\begin{align}
&a_{i_k} =
\begin{cases}
&\max\{e_{i_k}, \tau_{i_{k-1}}(b_{i_{k-1},i_k}) + t_{i_{k-1},i_k}\} + s_{i_k}, \quad \text{if}~i_k \in N \cup \{0\}\\
&\tau_{i_{k-1}}(b_{i_{k-1},i_k}) + t_{i_{k-1},i_k}, \quad \text{if}~i_k \in R
\end{cases}\\
\label{func.for}
&f_{i_k}(t) =
\begin{cases}
&f_{i_{k-1}}(\min\{t - s_{i_k} - t_{i_{k-1},i_k}, l_{i_{k-1}} + s_{i_{k-1}}\}) - b_{i_{k-1},i_k}, \quad \text{if}~i_k \in N \cup \{0\}\\
&\min\left\{\max_{x \in[\tau_{i_{k-1}}(b_{i_{k-1},i_k}), t-t_{i_{k-1},i_k}]}r_{i_k}\left(r^{-1}_{i_k}\left(f_{i_{k-1}}(x) - b_{i_{k-1},i_k}\right) + t - t_{i_{k-1},i_k} - x\right), B\right\}, \quad \text{if}~i_k \in R
\end{cases}\\
&=
\begin{cases}
&f_{i_{k-1}}(\min\{t - s_{i_k} - t_{i_{k-1},i_k}, l_{i_{k-1}} + s_{i_{k-1}}\}) - b_{i_{k-1},i_k}, \quad \text{if}~i_k \in N \cup \{0\}\\
&\min\left\{r_{i_k}(\hat{f}_{i_k}(t)  + t - t_{i_{k-1},i_k}), B\right\}, \quad \text{if}~i_k \in R,
\end{cases}
\end{align}
\end{small}
where $\hat{f}_{i_k}(t) = \max_{x \in[\tau_{i_{k-1}}(b_{i_{k-1},i_k}), t-t_{i_{k-1},i_k}]}\left\{r^{-1}_{i_k}\left(f_{i_{k-1}}(x) - b_{i_{k-1},i_k}\right) - x\right\}$, and initially $f_{i_0}(t) = B,~\forall~t \in [e_0, l_0]$.

\begin{figure}
\centering
\subfigure[The charging function]{
\label{fig.cf}
\begin{minipage}[b]{0.3\textwidth}
\includegraphics[width=1\textwidth]{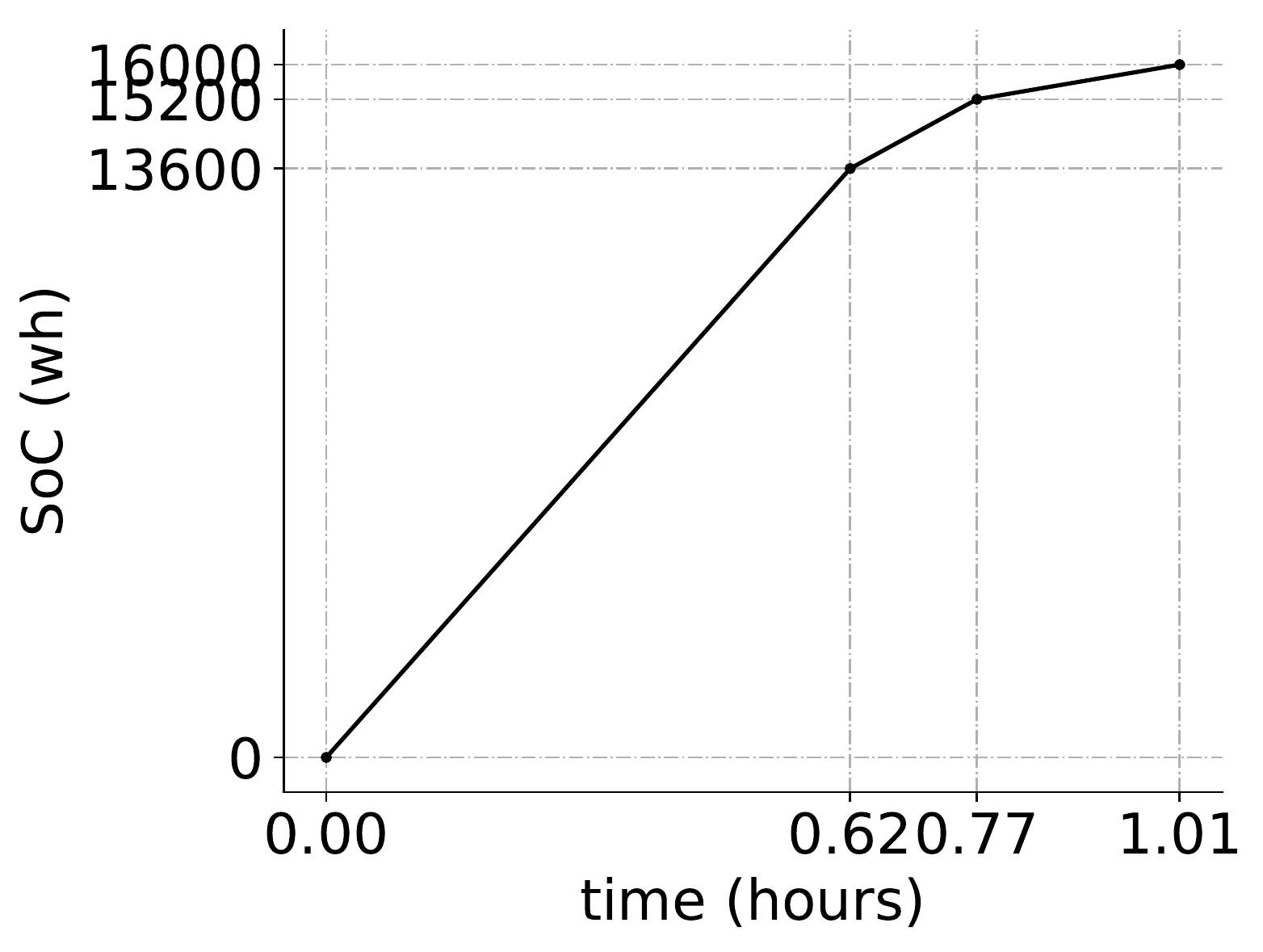} \\
\end{minipage}}
\subfigure[$f_j(t)$ when $j \in R$]{
\label{fig.fsta}
\begin{minipage}[b]{0.3\textwidth}
\includegraphics[width=1\textwidth]{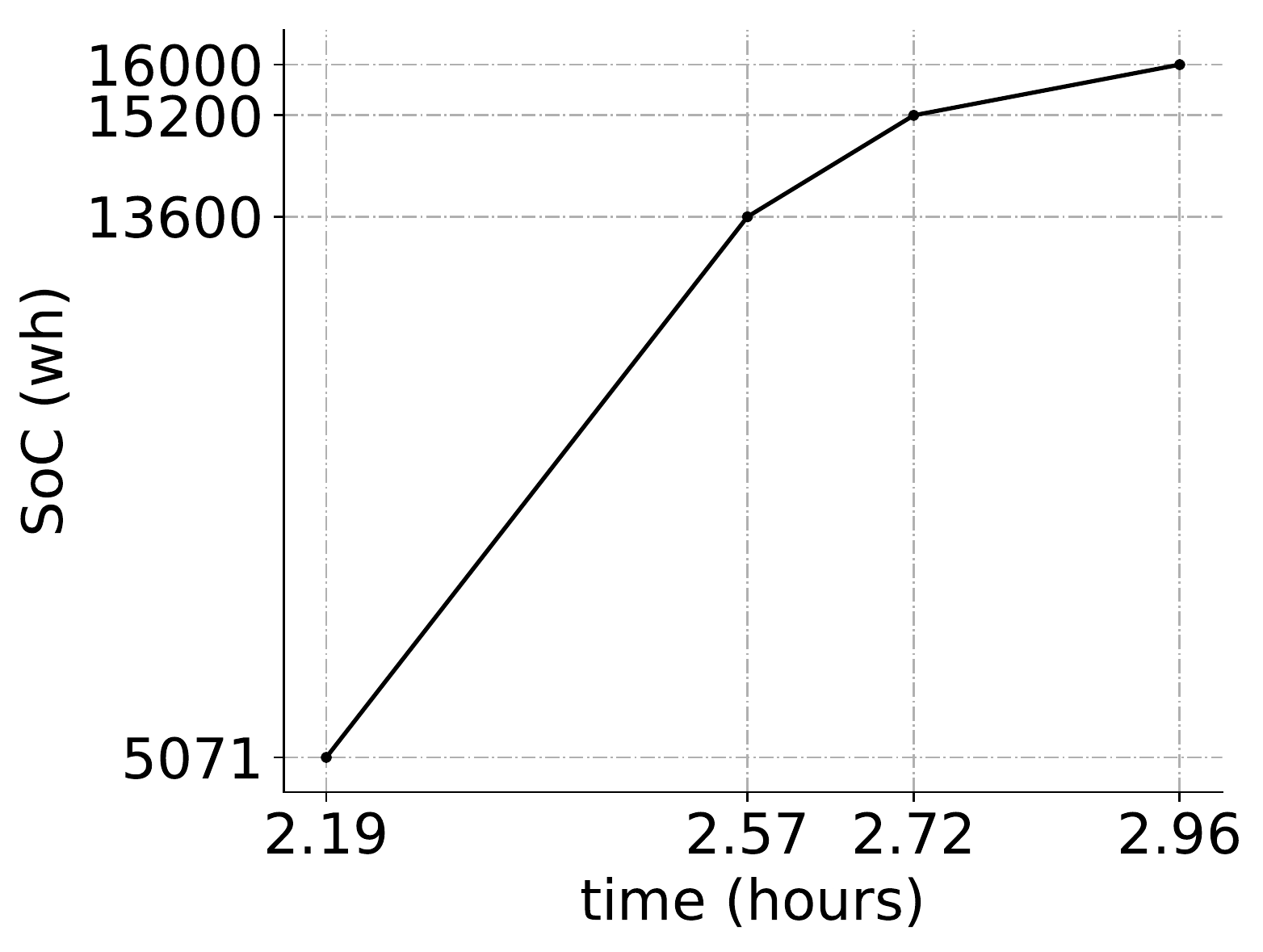}\\
\end{minipage}}
\subfigure[$f_j(t)$ when $j \in N$]{
\label{fig.fcust}
\begin{minipage}[b]{0.3\textwidth}
\includegraphics[width=1\textwidth]{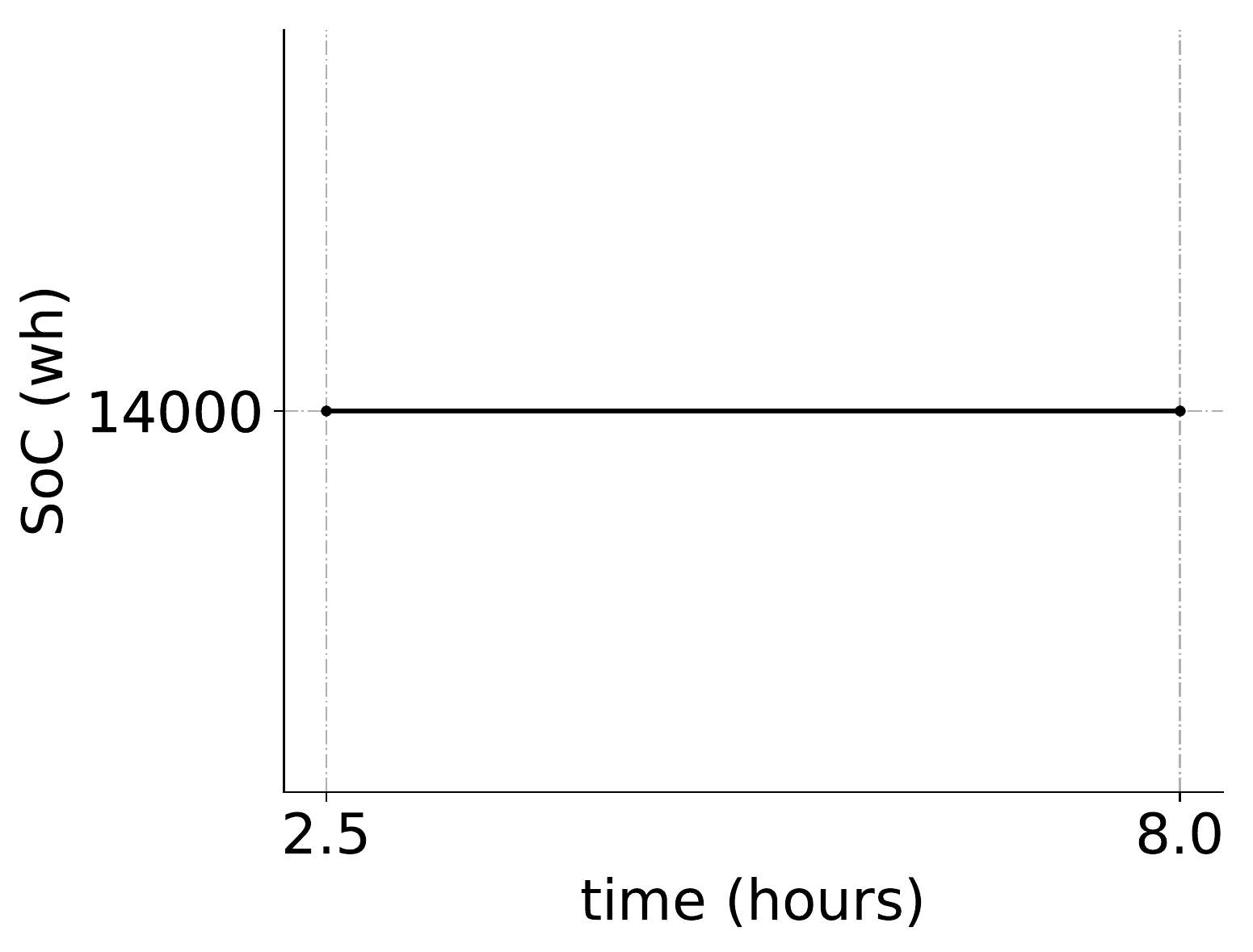} \\
\end{minipage}}
\label{fig.extension}
\caption{An example of the forward extension}
\end{figure}

In the sequel we present an illustrative example showing the forward extension. For simplicity, for each pair of nodes the travel time is set to 2h and the battery consumption is set to 2000wh. Furthermore, the battery capacity is 16000wh, the service time is 0.5h, the time window is $[2,7.5]$ for all customers and the depot's time window is set to $[0,10]$. The charging function is shown in Figure \ref{fig.cf} at the depot, and we have $f_0(t) = 16000, \forall t \in[0,10]$. When traversing arc $(0,j)$, we compute the function $f_j(t)$ using formula (\ref{func.for}). If $j$ is a charging station, the function $f_j(t)$ is depicted in Figure \ref{fig.fsta}. Otherwise, if $j \in N$, the function $f_j(t)$ is calculated as $f_j(t) = f_0(\min\{t - s_{i_j} - t_{0,j}, l_0 + s_0\})-2000=14000$wh, and is shown in Figure \ref{fig.fcust}.

\subsection{The backward recursive functions}
\label{subsec:Backrecursive}
For the backward search, we define a partial path $p = (i_k,i_{k-1},\ldots,i_0)$ where $i_0 = 0$. We denote $d_{i_k}$ the latest departure time at node $i_k$, and let $g_{i_k}(t)$ be the function representing the minimum battery level of the vehicle when leaving node $i_k$ at time $t$. This minimum battery level must ensure partial path $p$ is feasible. We further define $\rho_{i_k}(x) = \max\{t|g_{i_k}(t) \leq x\}$ as the function representing the latest departure time at $i_k$ that guarantees the battery level doesn't exceed the level $x$. We now elaborate on the recursive calculations for $d_{i_k}$ and $g_{i_k}(t)$, where we differentiate between two cases:

\begin{enumerate}[\text{Case} 1.]
\item Node $i_{k-1}$ is a customer or the depot\\
On the one hand, we know that the latest departure time at $i_k$ cannot be later than $l_{i_k}+s_{i_k}$ the sum of the upper bound of its time window and service time. On the other hand, as the maximum value of $g_{i_{k-1}}(t)$ is $B - b_{i_k,i_{k-1}}$, the latest departure time at $i_k$ can be calculated as $\rho_{i_{k-1}}(B - b_{i_k,i_{k-1}})- t_{i_k,i_{k-1}} - s_{i_{k-1}}$. Hence, $d_{i_k}=\min\{l_{i_k} + s_{i_k}, \rho_{i_{k-1}}(B - b_{i_k,i_{k-1}})- t_{i_k,i_{k-1}} - s_{i_{k-1}} \}$.\\
The minimum battery level $g_{i_k}(t)$ can be calculated as $g_{i_{k-1}}(t') +  b_{i_k,i_{k-1}}$, where $t'$ is the departure time from $i_{k-1}$.  The time $t'$ is computed as $t + t_{i_k,i_{k-1}}+s_{i_{k-1}}$, but can also not be earlier than $e_{i_{k-1}} + s_{i_{k-1}}$. Therefore, $g_{i_k}(t) = g_{i_{k-1}}(\max\{t + t_{i_k,i_{k-1}}+s_{i_{k-1}}, e_{i_{k-1}}+ s_{i_{k-1}}\}) + b_{i_k,i_{k-1}}$

\item Node $i_{k-1}$ is a CS\\
The departure time at $i_k$ is computed as the departure time $\rho_{i_{k-1}}(B - b_{i_k,i_{k-1}})$ from $i_{k-1}$ reduced by the travel time from $i_k$ to $i_{k-1}$ . So we have $d_{i_k}=\rho_{i_{k-1}}(B - b_{i_k,i_{k-1}}) - t_{i_k,i_{k-1}}$.\\
The minimum battery level $g_{i_k}(t)$ required at $i_k$ is equal to the battery level when leaving $i_{k-1}$ minus the amount of energy charged at $i_k$. If we let the departure time at $i_{k-1}$ be $t'$, the time $t'$ lies in the interval $[t + t_{i_k,i_{k-1}}, \rho_{i_{k-1}}(B - b_{i_k,i_{k-1}})]$. Moreover, the charging time is equal to $t' - t - t_{i_k,i_{k-1}}$. Hence, $g_{i_k}(t)$ is computed as $\min_{t' \in[t + t_{i_k,i_{k-1}}, \rho_{i_{k-1}}(B - b_{i_k,i_{k-1}})]}r_{i_{k-1}}\left(\max\left\{r^{-1}_{i_{k-1}}\left(g_{i_{k-1}}(t')\right) - t' + t + t_{i_k,i_{k-1}}, 0\right\}\right) + b_{i_k,i_{k-1}}$.
\end{enumerate}

In the sequel, we show how $g_{i_k}(t)$ and $d_{i_k}$ can be calculated recursively. We first prove the following lemma.

\begin{lemma}
\label{lemma:2}
The function $g_{i_k}(t)$ is a non-decreasing piecewise linear function.
\end{lemma}
\textit{Proof of Lemma~\ref{lemma:2}:}
Let $t_1, t_2$ denote two possible departure times at $i_k$ such that $t_2 \geq t_1$. We need to show that $g_{i_k}(t_2) \geq g_{i_k}(t_1)$, and do this by induction. First, we consider the second to last visited node $i_1$ along partial path $p$. If $i_1$ is a customer, then $g_{i_1}(t) = g_0(\max\{t + t_{i_k,i_0}, e_{i_0}\} + s_{i_0}) + b_{i_k,i_0} = b_{i_k,i_0}$, which is a constant. If $i_1$ is a CS, then $g_{i_1}(t)= r_{i_0}\left(\max\left\{\min_{x \in[t + t_{i_k,i_0}, \rho_{i_0}(B - b_{i_k,i_0})]}\left\{r^{-1}_{i_{k-1}}\left(g_{i_{k-1}}(x)\right) - x\right\} + t + t_{i_k,i_0}, 0\right\}\right) + b_{i_k,i_0}$  which is a non-decreasing piecewise linear function because the function $r_{i_k}(t)$ is a non-decreasing piecewise linear function.

We now assume $g_{i_{k-1}}(t)$ is a non-decreasing piecewise linear function, and differentiate between two cases.\\
First, if $i_k$ is a customer, we have,
\begin{align*}
g_{i_k}(t_2) - g_{i_k}(t_1) = g_{i_{k-1}}(\max\{t_2 + t_{i_k,i_{k-1}}, e_{i_{k-1}}\} + s_{i_{k-1}}) - g_{i_{k-1}}(\max\{t_1 + t_{i_k,i_{k-1}}, e_{i_{k-1}}\} + s_{i_{k-1}}) \geq 0.
\end{align*}
Second, If $i_k$ is a CS then,
\begin{align*}
g_{i_k}(t_2) - g_{i_k}(t_1)=
r_{i_{k-1}}\left(\max\left\{\hat{g}_{i_{k-1}}(t_2) + t + t_{i_k,i_{k-1}}, 0\right\}\right) - r_{i_{k-1}}\left(\max\left\{\hat{g}_{i_{k-1}}(t_1) + t + t_{i_k,i_{k-1}}, 0\right\}\right),
\end{align*}
where $\hat{g}_{i_{k-1}}(t) = \min_{x \in[t + t_{i_k,i_{k-1}}, \rho_{i_{k-1}}(B - b_{i_k,i_{k-1}})]}\left\{r^{-1}_{i_{k-1}}\left(g_{i_{k-1}}(x)\right) - x\right\}$.

Since $r_{i_{k-1}}(t)$ is a non-decreasing function, we only need to verify that $\hat{g}_{i_{k-1}}(t_2) \geq \hat{g}_{i_{k-1}}(t_1)$. We have:
\begin{equation*}
\begin{split}
\hat{g}_{i_{k-1}}(t_2) - \hat{g}_{i_{k-1}}(t_1)
&= \min_{x \in[t_2 + t_{i_k,i_{k-1}}, \rho_{i_{k-1}}(B - b_{i_k,i_{k-1}})]}\left\{r^{-1}_{i_{k-1}}\left(g_{i_{k-1}}(x)\right) - x\right\}\\
&-\min_{x \in[t_1 + t_{i_k,i_{k-1}}, \rho_{i_{k-1}}(B - b_{i_k,i_{k-1}})]}\left\{r^{-1}_{i_{k-1}}\left(g_{i_{k-1}}(x)\right) - x\right\}\\
&\geq 0,
\end{split}
\end{equation*}
as desired. $\square$

Now, $d_{i_k}$ and $g_{i_k}(t)$ can be computed according to the following recursive operations:
\begin{small}
\begin{align}
&d_{i_k} =
\begin{cases}
&\min\{l_{i_k} + s_{i_k}, \rho_{i_{k-1}}(B - b_{i_k,i_{k-1}})- t_{i_k,i_{k-1}} - s_{i_{k-1}} \}, \quad \text{if}~i_{k-1} \in N \cup \{0\}\\
&\rho_{i_{k-1}}(B - b_{i_k,i_{k-1}})- t_{i_k,i_{k-1}}, \quad \text{if}~i_{k-1} \in R
\end{cases}\\
\label{func.back}
&g_{i_k}(t) =
\begin{cases}
&g_{i_{k-1}}(\max\{t + t_{i_k,i_{k-1}}, e_{i_{k-1}}\} + s_{i_{k-1}}) + b_{i_k,i_{k-1}}, ~ \text{if}~i_{k-1} \in N \cup \{0\}\\
&\min_{x \in[t + t_{i_k,i_{k-1}}, \rho_{i_{k-1}}(B - b_{i_k,i_{k-1}})]}r_{i_{k-1}}\left(\max\left\{r^{-1}_{i_{k-1}}\left(g_{i_{k-1}}(x)\right) - x + t + t_{i_k,i_{k-1}}, 0\right\}\right) + b_{i_k,i_{k-1}}, ~ \text{if}~i_{k-1} \in R
\end{cases}\\
&=
\begin{cases}
&g_{i_{k-1}}(\max\{t + t_{i_k,i_{k-1}}, e_{i_{k-1}}\} + s_{i_{k-1}}) + b_{i_k,i_{k-1}}, ~ \text{if}~i_{k-1} \in N \cup \{0\}\\
&r_{i_{k-1}}\left(\max\left\{\hat{g}_{i_{k-1}}(t) + t + t_{i_k,i_{k-1}}, 0\right\}\right) + b_{i_k,i_{k-1}}, ~ \text{if}~i_{k-1} \in R,
\end{cases}
\end{align}
\end{small}
where $\hat{g}_{i_{k-1}}(t) = \min_{x \in[t + t_{i_k,i_{k-1}}, \rho_{i_{k-1}}(B - b_{i_k,i_{k-1}})]}\left\{r^{-1}_{i_{k-1}}\left(g_{i_{k-1}}(x)\right) - x\right\}$, and initially $g_{i_0}(t) = 0,~\forall~t \in [e_0, l_0]$.

\begin{figure}
\centering
\subfigure[$g_k(t)$,$k \in R$ or $k \in N$]{
\label{fig.bsta11}
\begin{minipage}[b]{0.3\textwidth}
\includegraphics[width=1\textwidth]{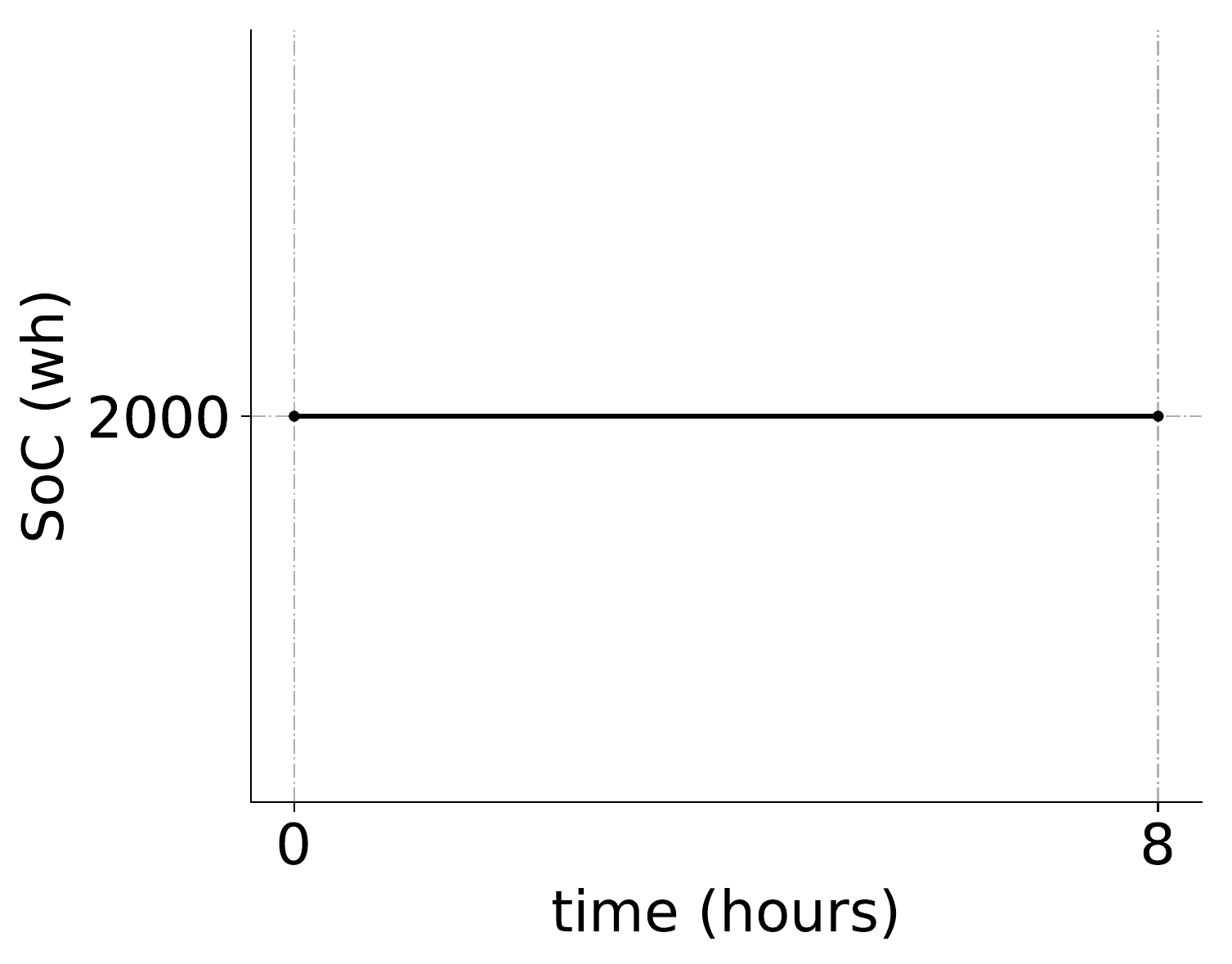}\\
\end{minipage}}
\subfigure[$g_c(t)$,$k \in R$]{
\label{fig.bsta}
\begin{minipage}[b]{0.36\textwidth}
\includegraphics[width=1\textwidth]{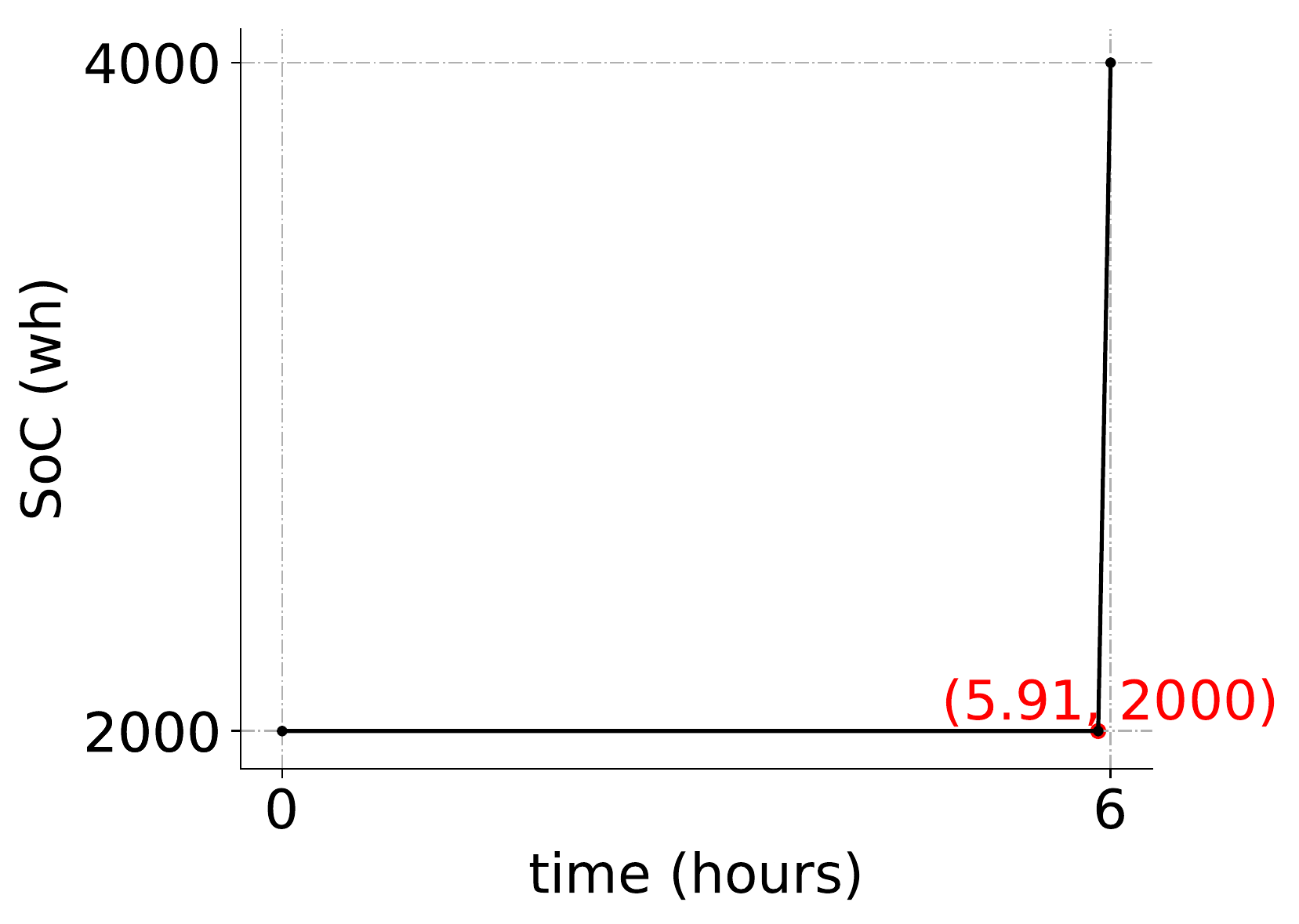}\\
\end{minipage}}
\subfigure[$g_c(t)$,$k \in N$]{
\label{fig.bcust}
\begin{minipage}[b]{0.3\textwidth}
\includegraphics[width=1\textwidth]{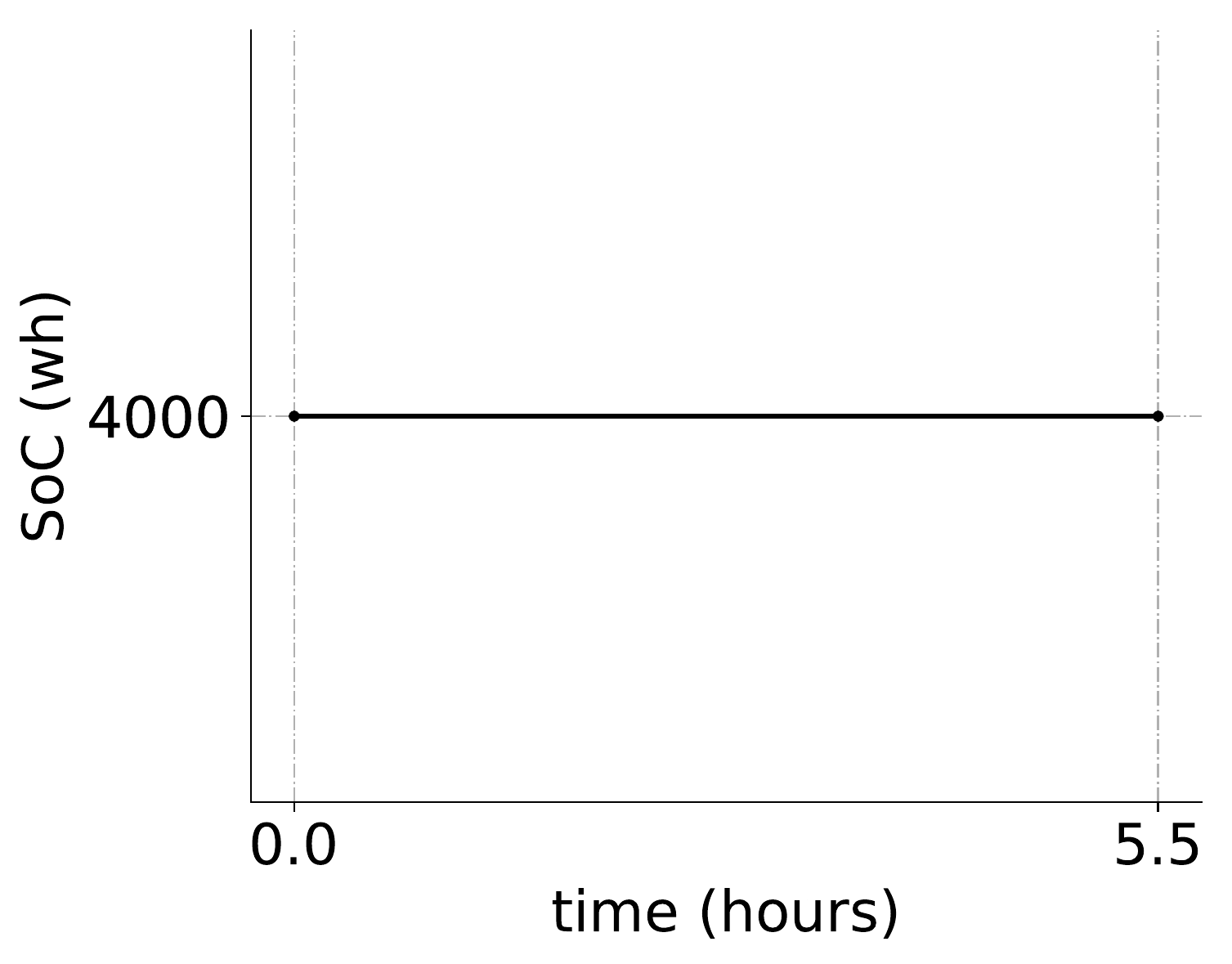}\\
\end{minipage}}
\label{fig.bextension}
\caption{An example of the backward extension}
\end{figure}

In the sequel, we present an illustrative example showing the backward extension. Using the same data as for the example on the forward extension, we further consider a partial path $(c,k,0)$. We have $g_0(t) = 0, \forall t \in [0,10]$. According to formula (\ref{func.back}), $g_k(t) = 2000, \forall t \in [0,8]$ regardless of whether $k$ is a customer or not, and is shown in Figure \ref{fig.bsta11}. Figures \ref{fig.bsta} and Figure \ref{fig.bcust} depict the function $g_c(t)$ in case $k$ is a charging station or a customer, respectively. Figure \ref{fig.bsta} shows that the minimum battery level required to finish this partial path grows rapidly after time 5.91h because the charging time required for the EV to charge from 2000wh to 4000wh is computed as 0.091h based on the charging function. Moreover, no charging is performed at $k$ when the EV leaves $c$ after time 6h, as only 4 hours are required to return back to the depot.

\subsection{The forward label-setting algorithm}
Let $L_i = (r_i, h_i, a_i, f_i(t), V_i)$ be a forward label representing a partial path $p(L_i)$ starting at the depot $0$ and ending in node $i$ where:
\begin{itemize}
    \item $r_i$ is the total collected duals along the partial path;
    \item $h_i$ is the total demand of the customers visited along the partial path;
    \item $a_i$ is the earliest arrival time at node $i$;
    \item $f_i(t)$ is the maximum battery level of the vehicle when it departs from node $i$ at time $t$;
    \item $V_i$ is the set of nodes visited along the partial path.
\end{itemize}

The basic operation in the label-setting algorithm consists of extending an existing label $L_i = (r_i, h_i, a_i, f_i(t), V_i)$ along an arc $(i, j) \in A$ to create a new label $L_j = (r_j, h_i, a_j, f_j(t), V_j)$ such that:
\begin{align}
&r_j =
\begin{cases}
&r_i + \mu_j, \quad \text{if}~ j \in N \cup \{0\}\\
&r_i, \quad \text{if}~ j \in R\\
\end{cases}\\
&h_j =
\begin{cases}
&h_i + q_j, \quad \text{if}~ j \in N \cup \{0\}\\
&h_i, \quad \text{if}~ j \in R\\
\end{cases}\\
&a_j =
\begin{cases}
\label{ext:a}
&\max\{e_j, \tau_{i}(b_{i,j}) + t_{i,j}\} + s_i, \quad \text{if}~ j \in N \cup \{0\}\\
&\tau_{i}(b_{i,j}) + t_{i,j}, \quad \text{if}~ j \in R\\
\end{cases}\\
&f_{j}(t) =
\begin{cases}
\label{ext:f}
&f_{i}(\min\{t-s_j-t_{i,j},l_i + s_i\})-b_{i,j}, \quad \text{if}~ j \in N \cup \{0\} \\
&\min\left\{\max_{x \in [\tau_{i}(b_{i,j}), t - t_{i,j}]}r_j\left(r^{-1}_{j}\left(f_{i}(x) - b_{i,j}\right) + t - t_{i,j} - x\right), B\right\}, \quad \text{if}~ j \in R \\
\end{cases}\\
&V_j = V_i  \cup \{j\}.
\end{align}

For a label $L_i$, let $P(L_i)$ denote the set of feasible extensions of partial path $p(L_i)$. An extension is feasible if it extends $p(L_i)$ all the way to the depot without violating time windows, storage capacity and battery level constraints. For a partial path $p^{\prime} \in P(L_i)$, let $p(L_i) \otimes p^{\prime}$ be the complete feasible route obtained by extending $p(L_i)$ with $p^{\prime}$, and $\bar{c}(p(L_i) \otimes p^{\prime})$ be its reduced cost.
To limit the combinatorial growth of labels in the label-setting algorithm, a dominance test is applied.
\begin{definition}
For any two labels $L^1_i$ and $L^2_i$ ending at the same node $i$, $L^1_i$ dominates $L^2_i$ if:
\begin{enumerate}
  \item $P(L^2_i) \subseteq P(L^1_i)$
  \item $\bar{c}(p(L^2_i) \otimes p^{\prime}) \geq \bar{c}(p(L^1_i) \otimes p^{\prime}), \quad \forall p^{\prime} \in P(L^2_i)$ \label{def:1.2}
\end{enumerate}
\label{def:1}
\end{definition}

Definition \ref{def:1} states that any feasible extension of $L^2_i$ is also feasible for $L^1_i$. Furthermore, extending $L^1_i$ must always result in a better route. However, it is impractical to check the conditions of Definition \ref{def:1} as all feasible extensions of $L^1_i$ and $L^2_i$ are required to be evaluated. Consequently, sufficient and tractable dominance rules are used. A set of intuitive dominance rules that can be applied to our label-setting algorithm is proposed in the following proposition:

\begin{proposition}[Dominance 1]
Given two labels $L^1_i = (r^1_i, h^1_i, a^1_i, f^1_i(t), V^1_i)$ and $L^2_i = (r^2_i, h^2_i, a^2_i, f^2_i(t), V^2_i))$ associated with the same node, $L^1_i$ dominates $L^2_i$ if:
\begin{align}
&r^1_i \geq r^2_i \label{wdom:1}\\
&h^1_i \leq h^2_i \label{wdom:2}\\
&a^1_i \leq a^2_i \label{wdom:3}\\
&V^1_i \subseteq V^2_i \label{wdom:4}\\
&f^1_i(t) \geq f^2_i(t), \quad \forall t \in [a^2_i + s_i, l_i + s_i] \label{wdom:5}
\end{align}
\label{pro:1}
\end{proposition}

Condition (\ref{wdom:1}) guarantees that for any $p^{\prime} \in P(L^2_i)$ the reduced cost of route $p(L^1_i) \otimes p^{\prime}$ is smaller than the reduced cost of route $p(L^2_i) \otimes p^{\prime}$, whereas conditions (\ref{wdom:2})-(\ref{wdom:5}) ensure that $P(L^2_i) \subseteq P(L^1_i)$.

\noindent \textit{Proof of Proposition~\ref{pro:1}:}
Consider two labels $L^1_i = (r^1_i, h^1_i, a^1_i, f^1_i(t), V^1_i)$ and $L^2_i = (r^2_i, h^2_i, a^2_i, f^2_i(t), V^2_i))$ that satisfy the conditions of Dominance 1. Due to condition (\ref{wdom:4}), if $p(L^1_i) \otimes p^{\prime}$ is elementary then $p(L^2_i) \otimes p^{\prime}$ will also be elementary. Moreover, as $h^2_i + \sum_{j \in p'} q_j - q_i \leq Q$, condition (\ref{wdom:2}) ensures that $h^1_i + \sum_{j \in p'} q_j- q_i\leq Q$, implying that storage capacity constraints are satisfied. Furthermore, due to condition (\ref{wdom:3}), any feasible arrival time at node $i$ in route $p(L^2_i) \otimes p^{\prime}$ is also feasible in route $p(L^1_i) \otimes p^{\prime}$. In other words, route $p(L^1_i) \otimes p^{\prime}$ respects customer time windows. Finally, condition (\ref{wdom:5}) ensures that the battery constraints are also satisfied along route $p(L^1_i) \otimes p^{\prime}$.

Now, we show that $\bar{c}(p(L^2_i) \otimes p^{\prime}) \geq \bar{c}(p(L^1_i) \otimes p^{\prime}), \forall p^{\prime} \in P(L^2_i)$. Let $c_{p'}$ denotes the duration of partial path $p^{\prime}$. We have:
\begin{align}
\bar{c}(p(L^1_i) \otimes p^{\prime}) = a^1_i + c_{p'} - r^{1}_i - \sum_{j \in p'}\mu_j + \mu_i\\
\bar{c}(p(L^2_i) \otimes p^{\prime}) = a^2_i + c_{p'} - r^{2}_i - \sum_{j \in p'}\mu_j + \mu_i
\end{align}
Hence,
\begin{align}
\bar{c}(p(L^2_i) \otimes p^{\prime}) - \bar{c}(p(L^1_i) \otimes p^{\prime}) = a^2_i - a^1_i + r^{1}_i - r^{2}_i \geq 0.
\end{align}
This completes the proof of Proposition~\ref{pro:1}.$\square$

Dominance \ref{def:1}, is not easy to check and most likely will not discard too many labels. In particular, condition (\ref{wdom:5}) is too restrictive as it requires that, for any time $t \in [a^2_i+s_i , l_i+s_i]$, the maximum battery level when leaving node $i$ is larger when using path $p(L_i^1)$ than when using path $p(L_i^2)$. To overcome this weakness, we develop a stronger dominance test. For a to be dominated label, we use a set of labels to dominate it instead of only one label as it is the case in Dominance 1. We introduce the new dominance test in the following definition:

\begin{definition}
For a label $L_i$ and a set of labels $\mathcal{L}_i$ associated with the same node $i$, $L_i$ is dominated by $\mathcal{L}_i$ if:
\begin{enumerate}
  \item $P(L_i) \subseteq \bigcup_{L^{\prime}_i \in \mathcal{L}_i} P(L^{\prime}_i)$ \label{def:2:1}
  \item $\bar{c}(p(L_i) \otimes p^{\prime}) \geq \bar{c}(p(L^{\prime}_i) \otimes p^{\prime}), \quad \forall p^{\prime} \in P(L_i), L^{\prime}_i \in \mathcal{L}_i$ \label{def:2.2}
\end{enumerate}
\label{def:2}
\end{definition}

Definition \ref{def:2} states that we can always find at least one label $L^{\prime}_i \in \mathcal{L}_i$, so that any feasible extension of $L_i$ is also feasible for $L^{\prime}_i$. Furthermore, for all $p^{\prime} \in P(L_i)$, each label in $\mathcal{L}_i$ results in a cheaper route. However, similar to Definition \ref{def:1}, performing dominance according to Definition \ref{def:2} is computationally expensive. Thus, a set of sufficient dominance conditions are formulated in the following proposition:
\begin{proposition}[Dominance 2]\label{pro:2}
Given a label $L_i = (r_i, h_i, a_i, f_i(t), V_i)$ and a set of labels $\mathcal{L}_i$ associated with the same node $i$, $L_i$ is dominated by $\mathcal{L}_i$ if, for all $L^{\prime}_i = (r^{\prime}_i, h^{\prime}_i, a^{\prime}_i, f^{\prime}_i(t), V^{\prime}_i) \in \mathcal{L}_i$,
\begin{align}
&r_i \leq r^{\prime}_i \label{sdom:1}\\
&h_i \geq h^{\prime}_i \label{sdom:2}\\
&a_i \geq a^{\prime}_i \label{sdom:3}\\
&V_i \supseteq V'_i \label{sdom:4}\\
&f_i(t) \leq \max_{L^{\prime}_i \in \mathcal{L}_i} f^{\prime}_i(t), \quad \forall t \in [a_i + s_i, l_i + s_i]\label{sdom:5}
\end{align}
We refer to Dominance 2 as set dominance.
\end{proposition}

\noindent \textit{Proof of Proposition~\ref{pro:2}:}
Consider label $L_i = (r_i, h_i, a_i, f_i(t), V_i)$ and a set $\mathcal{L}_i$ satisfying dominance conditions (\ref{sdom:1}) - (\ref{sdom:5}). First, we can show condition (\ref{def:2.2}) of Definition \ref{def:2} holds in the same way we have shown condition \ref{def:1.2} of Definition \ref{def:1} holds in Proposition \ref{pro:1}. Then, since conditions (\ref{sdom:1}) - (\ref{sdom:4}) are similar to conditions (\ref{wdom:1}) - (\ref{wdom:4}), we only focus on condition (\ref{sdom:5}). We assume that the energy consumption along route $p \in P(L_i)$ is $b_p$, we have:
\begin{align}
\max_{L^{\prime}_i \in \mathcal{L}_i} f^{\prime}_i(t) \geq f_i(t) \geq b_p, \quad \forall t \in [a_i + s_i, l_i + s_i],
\end{align}
which implies that for each $t \in [a_i + s_i, l_i + s_i]$, there always exists a label $L^{\prime\prime}_i \in \mathcal{L}_i$ with $f^{\prime\prime}_i(t) = \max_{L^{\prime}_i \in \mathcal{L}_i} f^{\prime}_i(t)$ that can be used to construct a feasible route $p(L^{\prime\prime}_i) \otimes p$ and therefore, condition (\ref{def:2:1}) of Definition \ref{def:2} holds. This completes the proof of Proposition~\ref{pro:2}.$\square$

In the following proposition, we show that set dominance is stronger than Dominance 1.
\begin{proposition}\label{pro:3}
Set dominance is stronger than Dominance \ref{def:1}.
\end{proposition}
\noindent \textit{Proof of Proposition~\ref{pro:3}:}
To prove proposition \ref{pro:3}, we show that a label that is dominated according to Dominance \ref{def:1} is also dominated according to set dominance. So, we let $L_i$ be a label that is dominated by another label $L^{\prime}_i$ according to Dominance \ref{def:1}. Obviously, by setting $\mathcal{L}_i = \{L^{\prime}_i\}$, label $L_i$ must also be dominated by the set $\mathcal{L}_i$ according to set dominance. This completes the proof of Proposition~\ref{pro:3}.$\square$

Clearly, the strength of set dominance depends on the set $\mathcal{L}_i$. We choose the set $\mathcal{L}_i$ such that it consists of labels that dominate $L_i$ in all resources except the battery level related resource (i.e. $f_i(t)$) as $\mathcal{L}_i = \{L^{\prime}_i = (r^{\prime}_i, h^{\prime}_i, a^{\prime}_i, f^{\prime}_i(t), V^{\prime}_i)|r^{\prime}_i \geq r_i, h^{\prime}_i \leq h_i, a^{\prime}_i \leq a_i, V^{\prime}_i \subseteq V_i\}$. Increasing the size of $\mathcal{L}_i$ strengthens condition (\ref{sdom:5}) and thus the set dominance. Inspired by \cite{luo2016branch}, we implement the set dominance using a procedure based on an acyclic directed graph named {\em dominance graph}. Let $\mathbb{G}_i = (\mathbb{N}_i, \mathbb{A}_i)$ be the dominance graph related to node $i \in V$ where a node $u \in \mathbb{N}_i$ is associated with four attributes $r_u, h_u, a_u$ and $V_u$ that have the same meaning as for the label definition, and a set of non-dominated labels $\mathcal{L}_u$ ending in node $i$. For any two nodes $u,v \in \mathbb{N}_i$, there exists an arc from node $u$ to node $v$ in $\mathbb{A}_i$ if $r_u \geq r_v, h_u \leq h_v, a_u \leq a_v, V_u \subseteq V_v$ and at least one of the four inequalities is strict. For a node $v \in \mathbb{N}_i$, let $\mathbb{N}_i(v) = \{u \in \mathbb{N}_i|(u,v) \in \mathbb{A}_i\}$ be the set of nodes connected to $v$ by an arc $(u,v)$. We define the function $\mathbb{F}_v(t)$ as:
\begin{align}
\mathbb{F}_v(t) = \max \{\max_{L_i\in \mathcal{L}_v} f_i(t), \max_{u \in \mathbb{N}_i(v)} \mathbb{F}_u(t)\}.
\end{align}

\begin{figure}[htb]
\begin{center}
\includegraphics[scale=0.8]{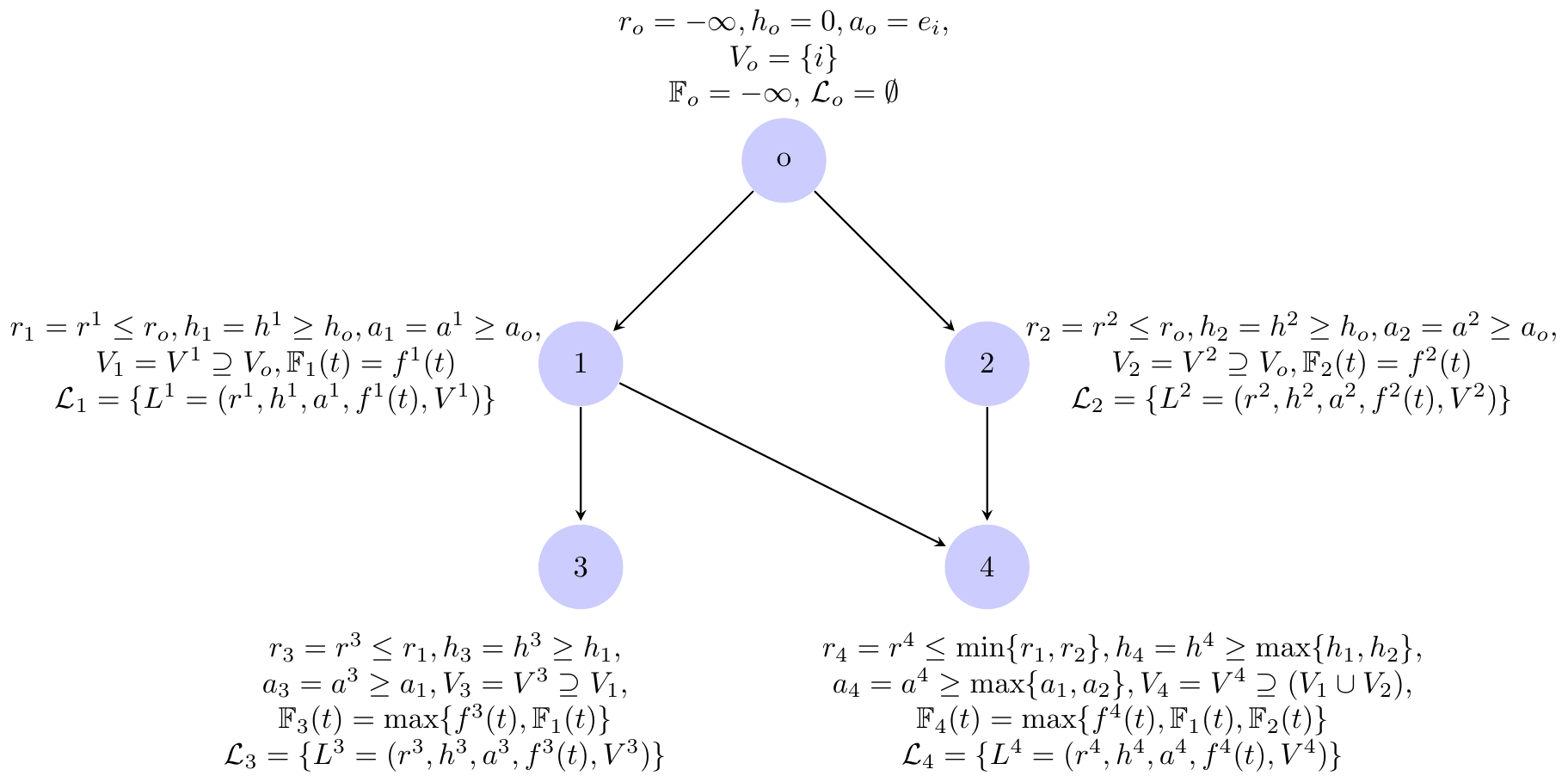}
\end{center}
\caption{An example of the dominance graph}
\label{fig.dg}
\end{figure}

An illustration of a dominance graph is presented in Figure \ref{fig.dg}. Now, we can formulate the following proposition for set dominance.
\begin{proposition}\label{pro:4}
A label $L_i = (r_i, h_i, a_i, f_i(t), V_i)$ is dominated by the dominance graph $\mathbb{G}_i$ if there exists a node $v \in \mathbb{N}_i$, such that:
\begin{align}
&r_i \leq r_v \label{gdom:1}\\
&h_i \geq h_v \label{gdom:2}\\
&a_i \geq a_v \label{gdom:3}\\
&V_i \supseteq V_v \label{gdom:4}\\
&f_i(t) \leq \mathbb{F}_v(t),\quad \forall t \in [a_i + s_i, l_i + s_i] \label{gdom:5}
\end{align}
\end{proposition}

\noindent \textit{Proof of Proposition~\ref{pro:4}:}
Consider a label $L_i = (r_i, h_i, a_i, f_i(t), V_i)$ and a node $v \in \mathbb{N}_i$ satisfying dominance rules (\ref{gdom:1}) - (\ref{gdom:5}). First, we can show condition (\ref{def:2.2}) of Definition \ref{def:2} holds in the same way we have shown condition \ref{def:1.2} of Definition \ref{def:1} holds in Proposition \ref{pro:1}. Then, since conditions (\ref{gdom:1}) - (\ref{gdom:4}) are similar to conditions (\ref{wdom:1}) - (\ref{wdom:4}), we only focus on condition (\ref{gdom:5}). We assume that the energy consumption along route $p \in P(L_i)$ is $b_p$. We have:
\begin{align}
\mathbb{F}_v(t) = \max \{\max_{L'_i\in \mathcal{L}_v} f'_i(t), \max_{u \in \mathbb{N}_i(v)} \mathbb{F}_u(t)\} \geq f_i(t) \geq b_p, \quad \forall t \in [a_i + s_i, l_i + s_i],
\end{align}
which implies that for each $t \in [a_i + s_i, l_i + s_i]$, there always exists a label $L''_i \in \mathcal{L}_v \cup \mathcal{L}_{u \in \mathbb{N}_i(v)}$ with $f''_i(t) \geq b_p$ that can be used to construct a feasible route $p(L''_i) \otimes p$ and therefore, condition (\ref{def:2:1}) of Definition \ref{def:2} holds. This completes the proof of Proposition~\ref{pro:4}.$\square$

We initially create a dummy root node $o \in \mathbb{N}_i$ with $r_o = -\infty, h_o = 0, a_o = e_i, V_o = \{i\}$ and $\mathbb{F}_o = -\infty, \forall t \in [e_i, l_i]$. In order to dominate label $L_i$, we explore the nodes of dominance graph $\mathbb{G}_i$. If conditions (\ref{gdom:1}) - (\ref{gdom:5}) are satisfied, label $L_i$ is dominated. Otherwise, label $L_i$ is used to update the dominance graph $\mathbb{G}_i$. If a node $v$ with the same attributes as label $L_i$ is found, label $L_i$ is added to the set $\mathcal{L}_v$, and the function $f_i(t)$ is used to update the function $\mathbb{F}_v(t)$. If no such node exists, a new node $v^{\prime}$ with the same attributes as label $L_i$ is created and added to the dominance graph $\mathbb{G}_i$. The function $\mathbb{F}_{v^{\prime}}(t)$ of the new node $v^{\prime}$ is initialized as $f_i(t)$. Besides, $f_i(t)$ is also used to update the function $\mathbb{F}_u(t), \forall u \in N_{i}(v^{\prime})$. If a label $L_i$ has no contribution to the function $\mathbb{F}_u(t)$, it is discarded as it obviously will be dominated after new labels are added. Clearly, the leaf nodes (i.e. nodes without outgoing arcs) in the graph have a larger $\mathbb{F}_u(t)$ function, and are therefore explored first as they are more likely to result in a successful dominance. A description of the implementation of the set dominance rules is presented in Algorithm \ref{alg:al}.

\begin{algorithm}[htp]
\caption{{\em AddLabel($\mathbb{G}_i$, $L_i$)}} \label{alg:al}
\begin{small}
\begin{algorithmic}[1]
\STATE Conduct dominance test between label $L_i$ and the leaf nodes in $\mathbb{G}_i$;
\IF{Label $L_i$ is dominated by a leaf node}
  \STATE Return \textbf{False};
\ELSE
  \STATE Explore graph $\mathbb{G}_i$ by the depth-first-search strategy;
  \STATE Construct the set of nodes $\mathbb{G}^{\prime}_i \subset \mathbb{G}_i$. Conditions (\ref{sdom:1})--(\ref{sdom:5}) are satisfied by each node $u \in \mathbb{G^{\prime}}_i$ but not by any successive nodes of $u$;
  \STATE Conduct dominance tests between label $L_i$ and the nodes in $\mathbb{G}^{\prime}_i$;
  \IF{Label $L_i$ is dominated by a node in $\mathbb{G}^{\prime}_i$}
    \STATE Return \textbf{False};
  \ELSE
    \STATE Search for a node $u \in \mathbb{G}_i$ with $r_u = r_i$, $h_u = h_i$, $a_u = a_i$ and $V_u = V_i$;
    \IF{Node $u$ exists}
      \STATE Add $L_i$ to $\mathcal{L}_u$ and set $\mathbb{F}_u(t) := \max\{\mathbb{F}_u(t), f_i(t)\}$
    \ELSE
      \STATE Create a new node $u^{\prime}$ with $r_{u^{\prime}} = r_i$, $h_{u^{\prime}} = h_i$, $a_{u^{\prime}} = a_i$, $V_{u^{\prime}} = V_i$, $\mathbb{F}_{u^{\prime}}(t) = f_i(t)$ and $\mathcal{L}_{u^{\prime}}=\{L_i\}$;
      \STATE Add node $u^{\prime}$ to graph $\mathbb{G}_i$ and set $u := u^{\prime}$;
    \ENDIF
    \STATE Select the set of nodes $\mathbb{G}^{\prime\prime}_i \subset \mathbb{G}_i$ which have at least a path to node $u$;
    \FOR{Each node $u^{\prime\prime} \in \mathbb{G}^{\prime\prime}_i$}
      \STATE Set $\mathbb{F}_{u^{\prime\prime}}(t):=\max\{\mathbb{F}_{u^{\prime\prime}}(t), f_u(t)\}$;
      \STATE Remove labels in $\mathcal{L}_{u^{\prime\prime}}$ which have no contribution to $\mathbb{F}_{u^{\prime\prime}}(t)$;
      \IF{$\mathcal{L}_{u^{\prime\prime}} = \emptyset$}
        \STATE Remove node $u^{\prime\prime}$ from $\mathbb{G}_i$;
      \ENDIF{}
    \ENDFOR
  \ENDIF
 \ENDIF
\STATE Return \textbf{True};
\end{algorithmic}
\end{small}
\end{algorithm}

A description of the label-setting algorithm is given in Algorithm \ref{alg:ls}. The label-setting algorithm maintains two pools for each node $i \in V$: pool $\Pi_i$ to store labels that have not been extended and pool $\Phi_i$ to store labels that have been extended. First, the two label pools and the dominance graphs are initialized, and an initial label $L_0$ associated with the depot node is created and added to $\Pi_0$. As long as $\Pi_i$ is non-empty, a label $L_i$ is extracted from $\Pi_i$, added to $\Phi_i$ and then extended along arc $(i,j)$ to create a new label $L_j$. If the newly created label $L_j$ survives the dominance test, it is added to pool $\Pi_j$. The pool $\Pi_j$ is reduced by removing all the labels dominated by $L_j$. 

\begin{algorithm}[htp]
\caption{{\em Label-Setting Algorithm}} \label{alg:ls}
\begin{small}
\begin{algorithmic}[1]
\STATE Create an initial label $L_0 = (-\infty, 0, e_0, f_0(t), V)$ where $f_0(t) = B~ \forall t \in [e_0, l_0]$;
\STATE Set $\Pi_0:=\{L_0\}$, $\Pi_i := \emptyset~\forall~i \in V_c$ and $\Phi_i:=\emptyset~\forall~i \in V$;
\STATE Create a dominance graph $\mathbb{G}_i~\forall~i \in V$;
\WHILE{$\cup_{i \in V}\Pi_i \neq \emptyset$}
  \STATE Select a label $L_i \in \Pi_i$ $(\Pi_i \neq \emptyset)$;
  \STATE Set $\Phi_i:=\Phi_i \cup \{L_i\}$;
  \FOR{Each arc $(i,j) \in A$}
    \IF{$j \in V_i$}
      \STATE Extend label $L_i$ through arc $(i,j)$ to create a new label $L_j$;
      \IF{AddLabel($\mathbb{G}_j$, $L_j$) = \textbf{True}}
        \STATE Set $\Pi_j:= (\Pi_j \cup \{L_j\} ) \cap \cup_{u \in \mathbb{G}_j}\mathcal{L}_u $;
      \ENDIF
    \ENDIF
  \ENDFOR
\ENDWHILE
\STATE Return the route which corresponds to the label in $\Phi_0$ with the minimum reduced cost.
\end{algorithmic}
\end{small}
\end{algorithm}

\subsection{The backward label-setting algorithm}
Let $B_i = (\tilde{r}_i, \tilde{h}_i, \tilde{a}_i, \tilde{g}_{i}(t), \tilde{V}_i)$ be a backward label representing a path from node $i$ to the depot $0$ where:
\begin{itemize}
  \item $\tilde{r}_i$ is the total collected duals along the partial path;
  \item $\tilde{h}_i$ is the total demand of the customers visited along the partial path;
  \item $\tilde{a}_i$ is the latest departure time from node $i$;
  \item $\tilde{g}_{i}(t)$ is the minimum battery level of the vehicle when it departs from node $i$ at time $t$;
 \item $\tilde{V}_i$ is the set of nodes visited along the partial path.
\end{itemize}
The basic operation in the backward label-setting algorithm consists of extending an existing label $B_i = (\tilde{r}_i, \tilde{h}_i, \tilde{a}_i, \tilde{g}_{i}(t), \tilde{V}_i)$ along an arc $(j, i) \in A$ to create a new label $B_j = (\tilde{r}_j, \tilde{h}_j, \tilde{a}_j, \tilde{g}_{j}(t), \tilde{V}_j)$ such that:
\begin{align}
&\tilde{r}_j =
\begin{cases}
&\tilde{r}_i + \mu_j, \quad \text{if}~ j \in N \cup \{0\}\\
&\tilde{r}_i, \quad \text{if}~ j \in R\\
\end{cases}\\
&\tilde{h}_i =
\begin{cases}
&\tilde{h}_i + q_j, \quad \text{if}~ j \in N \cup \{0\}\\
&\tilde{h}_i, \quad \text{if}~ j \in R\\
\end{cases}\\
&\tilde{a}_j =
\begin{cases}
&\min\{l_j + s_j, \tilde{\rho}_{i}(B - b_{j,i}) - t_{j,i} - s_i\}, \quad \text{if}~ i \in N \cup \{0\}\\
&\tilde{\rho}_{i}(B - b_{j,i}) - t_{j,i}, \quad \text{if}~ i \in R\\
\end{cases}\\
&\tilde{g}_{j}(t) =
\begin{cases}
&\tilde{g}_{i}(\max\{t + t_{j,i},e_i\} + s_i) + b_{j,i}, \quad \text{if}~ i \in N \cup \{0\} \\
&\min_{x \in [t + t_{j,i}, \tilde{\rho}_{i}(B - b_{j,i})]}r_i\left(\max \left\{r^{-1}_{i}\left(\tilde{g}_{i}(x)\right) - x + t + t_{j,i},0\right\}\right) + b_{j,i}, \quad \text{if}~ i \in R \\
\end{cases}\\
&\tilde{V}_j = \tilde{V}_i \cup \{j\}
\end{align}

The set dominance for the backward label-setting algorithm is presented in Proposition \ref{pro:5}. The proof of Proposition \ref{pro:5} is similar to the prove of Proposition \ref{pro:2}, and is therefore omitted.
\begin{proposition}[Dominance 3]\label{pro:5}
Given a label $B_i = (\tilde{r}_i, \tilde{h}_i, \tilde{a}_i, \tilde{g}_i(t), \tilde{V}_i)$ and a set of labels $\mathcal{B}_i$ associated with the same node $i$, $B_i$ is dominated by $\mathcal{B}_i$ if, for all $B^{\prime}_i = (r^{\prime}_i, h^{\prime}_i, a^{\prime}_i, f^{\prime}_i(t), V^{\prime}_i) \in \mathcal{B}_i$,
\begin{align}
&\tilde{r}_i \leq \tilde{r}^{\prime}_i \label{bsdom:1}\\
&\tilde{h}_i \geq \tilde{h}^{\prime}_i \label{bsdom:2}\\
&\tilde{a}_i \leq \tilde{a}^{\prime}_i \label{bsdom:3}\\
&\tilde{V}_i \supseteq \tilde{V}^{\prime}_i \label{bsdom:4}\\
&\tilde{g}_i(t) \geq \min_{B^{\prime}_i \in \mathcal{B}_i} \tilde{g}^{\prime}_i(t), \quad \forall t \in [a_i + s_i, l_i + s_i]\label{bsdom:5}
\end{align}
\end{proposition}

\subsection{Merging forward and backward labels}
The forward labels and backward labels are merged to construct complete routes. We use time as the bounding resource to stop forward and backward extensions. We denote $T^f$ and $T^b$ as the stopping thresholds for the forward and backward extensions, respectively. The value of $T^f$ and $T^b$ is not necessary half of the depot's time window as also pointed out by \citet{tilk2017asymmetry}. Therefore, we dynamically update the values of $T^f$ and $T^b$ during the search process. Firstly, we divide the time window of the depot into 16 slots, and set $T^f = e_0 + \frac{l_0 - e_0}{16}$ and $T^b = l_0 - \frac{l_0 - e_0}{16}$. The number of forward and backward labels generated during the search process is recorded. When the number of forward labels exceeds the number of backward labels, we subtract $\frac{l_0 - e_0}{16}$ from $T^b$. Otherwise, we set $T^f := T^f + \frac{l_0 - e_0}{16}$.

A forward label $L_i = (r_i, h_i, a_i, f_i(t), V_i)$ and a backward label $B_i = (\tilde{r}_i, \tilde{h}_i, \tilde{a}_i, \tilde{g}_{i}(t), \tilde{V}_i)$ associated with the same node $i$ can be joined together if they satisfy the following conditions:
\begin{align}
&h_i + \tilde{h}_i - q_i \leq Q, \label{join:1}\\
&a_i + s_i \leq \tilde{a}_i, \label{join:2}\\
&V_i \cap \tilde{V}_i = \emptyset, \label{join:3}\\
&\max f_{i}(t) \geq \min \tilde{g}_{i}(t), \quad t \in [a_i + s_i, \tilde{a}_i] \label{join:4}
\end{align}

Conditions (\ref{join:1}), (\ref{join:2}) and (\ref{join:4}) ensure storage vehicle capacity constraints, customer time windows and battery levels are respected, while condition (\ref{join:3}) guarantee that every customer is visited at most once. The reduced cost of the resulting complete route is
\begin{align}
&T - \max_{a_i \leq t_1 \leq t_2 \leq \tilde{a}_i, f_i(t_1) \geq \tilde{g}_i(t_2)}(t_2 - t_1) - r_i - \tilde{r}_i + \mu_i.
\end{align}

\section{Implementation Features}\label{sec:implementation}
In this section we present the features that we included in our exact algorithm to speed up the solution time.

\subsection{Bound computation}
To stop label extension early in the search process, we apply the $q$-route relaxation introduced by \citet{christofides1981exact} to compute a lower bound on the reduced cost of a partial path when extended all the way to the depot. This bounding rule is applied to both the forward and backward search. We further forbid cycles of two nodes to improve the obtained lower bounds. Using these lower bounds facilitates pruning labels that can not form a complete route with a negative reduced cost.

\subsection{Ng-route relaxation}
In order to accelerate the label-setting algorithm, we adopt the iterative method based on ng-route relaxation proposed by \citet{martinelli2014efficient}. Ng-routes are first introduced by \citet{baldacci2011new}. Each node $k$ is associated a {\em ng-set}, which can be composed of the set of closest nodes to it. If an {\em ng-route} has visited a node $k$, then it cannot be extended to $k$ again if $k$ exists in the intersections of ng-sets of those nodes visited after $k$. At the beginning, each node is associated with a small-sized ng-set. Afterwards, optimal ng-routes are obtained by invoking the label-setting algorithm. If the optimal ng-routes contain cycles, the label-setting algorithm will be invoked again with larger ng-sets of those nodes belonging to the cycles. This process terminates once the optimal ng-routes are elementary.

\subsection{Subset-row inequalities}
\label{subsec:sr}
To improve the quality of the obtained lower bounds at each branch-and-bound node, and hence find integer solutions faster, we adopt the subset-row inequality first introduced by \citet{jepsen2008subset}. In line with the literature, we consider subset-row inequalities defined by node sets $S$ of three nodes. In this case, at most one route in a feasible solution covers more than one node from the node set, and it can be expressed as:
\begin{align}
\sum_{p \in \Omega} \left \lfloor \sum_{p \in S} \frac{\alpha_{i,p}}{2} \right \rfloor \leq 1, \quad \forall~   S \subseteq \Omega , |S| = 3.
\end{align}
The separation of violated subset-row inequalities is done by full enumeration.

\subsection{Branching strategies}
Two branching rules are hierarchically performed in our algorithm to guarantee an integer optimal solution. Let $\bar{\theta}$ be the current solution of the LP relaxation of the set partitioning model. The following branching rules are performed:
\subsubsection*{Branching on the number of vehicles:}
Let $\bar{k}$ be the number of used vehicles which is equal to $\sum_{p \in \Omega}\bar{\theta}_{p}$. If $\bar{k}$ is fractional, we branch on the value of $\bar{k}$ to generate two child nodes where constraints $\sum_{p \in \Omega}\theta_{p} \leq \left \lfloor \bar{k} \right \rfloor$ and $\sum_{p \in \Omega}\theta_{p} \geq \left \lceil \bar{k} \right \rceil$ are added to their corresponding master problems, respectively. If $\bar{k}$ is integer, we branch on arcs.
\subsubsection*{Branching on arcs:}
Let $\gamma_{ijp}$ denote the number of times arc $(i,j)$ is traversed along route $p$. Then $\bar{z}_{ij} = \sum_{p \in \Omega} \gamma_{ijp}\bar{\theta}_{p} $ is the number of vehicles traversing arc $(i,j)$ in the current solution. Strong branching is applied as follows: for each arc $(i,j)$ with a fractional $\bar{z}_{ij}$, we create two child nodes in which constraints $\sum_{p \in \Omega} \gamma_{ijp}\theta_{p} \geq \left \lceil \bar{z}_{ij} \right \rceil$ and $\sum_{p \in \Omega} \gamma_{ijp}\theta_{p} \leq \left \lfloor \bar{z}_{ij} \right \rfloor$ are imposed in their corresponding master problems, respectively. For each arc we calculate the sum of the lower bounds obtained from the corresponding child nodes and select the arc with the largest value for branching.

\section{The Tabu Search Heuristic}
\label{sec:heuristic}
In order to solve larger instances faster, we develop a tabu search based heuristic. Tabu search based heuristics have proven to efficiently solve a variety of vehicle routing problems \citep{glover1989tabu}. In the following, we elaborate on the different features of our tabu search heuristic. In particular, we explain how we deal with charging related decisions. A pseudo-code of the developed tabu search is illustrated in Algorithm \ref{alg:tb1}.
\subsection{Initial solution}
At the beginning, the incumbent solution consists of a set of empty routes. The customer with the smallest insertion cost is added to the incumbent solution until all customers are served. If an unserved customer cannot be inserted due to battery capacity, we generate a set of routes $s = \{(0,cs,0),\forall cs \in R\}$ and try to include the unserved customers starting with the ones with the least insertion cost. If we still fail to insert all unserved customers, we create a route set $s' = \{(0,cs,cs,0),\forall cs \in R\}$ in which we again try to include the unserved customers. This process is repeated by increasing the number of charging stations in the new created routes until all customers are included in a route. Finally, the routes in which no customer is visited are removed. This procedure corresponds to line \ref{alg.ts.1} in Algorithm \ref{alg:tb1}.
\subsection{Neighborhood structure}
The Neighbor solutions are constructed by performing three different moves between two different routes in the incumbent solution: (1) relocate a customer to another position ({\em relocate}), (2) switch two arcs between two routes ({\em cross}), and (3) exchange the positions of two customers ({\em exchange}). The move that is not tabu and that results in the highest cost saving is selected in each iteration. Moreover, a move is only allowed if it results in a new global solution. This procedure corresponds to line \ref{alg.ts.5} and \ref{alg.ts.11} in Algorithm \ref{alg:tb1}.
\subsection{Feasibility check}
During the initial solution construction and neighbourhood search, a feasibility check is performed. A feasible route must respect vehicle storage capacity, customer time windows and battery capacity.  Therefore, after performing a move, the feasibility of the resulting routes is evaluated. This is done by, for each node $i_k$ on a route $r$, keeping track of the resources $a_{i_k}$, $f_{i_k}(t)$, $d_{i_k}$ and $g_{i_k}(t')$ introduced in section \ref{subsec:recursive} as well as the load $q_r$ on route $r$. If customer $i_a^*$ is to be inserted in route $r$, we calculate the value of $a_{i_a^*}$, $f_{i_a^*}(t)$, $d_{i_a^*}$ and $g_{i_a^*}(t)$ using the recursive functions introduced in section \ref{subsec:recursive} and ensure that:
\begin{align}
&q_r + q_{i_a^*} < Q,\\
&a_{i_a^*} \leq d_{i_a^*},\\
&\max f_{i_a^*}(t) \geq \min g_{i_a^*}(t'), \forall t \in[t_1,t_2], t' \in [t_3,t_4]
\end{align}
\subsection{Optimization of charging decisions}
Contrary to the exact algorithm, in the tabu search, the charging schedules are determined independently for a fixed sequence of visited customers on a route. The label-setting algorithm described in Algorithm \ref{alg:ls} is modified and applied to each route in the incumbent solution. For each route $r$, let $(i_0,...,i_k,i_{k+1},...,i_n)$ represents the sequence of customers and $V'$ denotes the set of customers visited in this sequence. Let $A_{i_k}'=\{(i_k,i_{k+1})\cup(i_k,j,i_{k+1}), \forall j \in R\}$ be the set of arcs going out of $i_k\in V'$ and $L'_{i_k} = (a'_{i_k}, b'_{i_k}, f'_{i_k}(t))$ be a label representing a path from the depot $0$ to node ${i_k}$ in which $a'_{i_k}$ is the earliest departure time at node ${i_k}$, $b'_{i_k}$ denotes the visited CS before node ${i_k}$ and $f'_{i_k}(t)$ is the maximum battery level of the vehicle when it departs from node ${i_k}$ at time $t$. The initial label is defined as $L'_0 = (e_0, -1, f_0(t))$, where $f_0(t) = B$, for all $t \in [e_0, l_0]$.

Label $L'_{i_k} = (a'_{i_k}, b'_{i_k}, f'_{i_k}(t))$ associated with node $i_k$ is extended along an arc $a \in A_{i_k}'$ to create a new label $L'_{i_{k+1}} = (a'_{i_{k+1}}, b'_{i_{k+1}}, f'_{i_{k+1}}(t))$ according to (\ref{ext:a}) and (\ref{ext:f}), and
\begin{align}
&b'_{i_{k+1}} =
\begin{cases}
&-1, \quad \text{if}~ a \in (i_k,i_{k+1})\\
&j, \quad \text{if}~ a \in (i_k,j,i_{k+1}),j \in R.\\
\end{cases}
\end{align}
The dominance rules from Proposition \ref{pro:2} are modified. Given a label $L'_{i_k} = (a'_{i_k}, b'_{i_k}, f'_{i_k}(t))$ and a set of labels $\mathcal{L}_i$ associated with the same node $i_k$, $L'_{i_k}$ is dominated by $\mathcal{L}_{i_k}$ if, for all $L^*_{i_k} = (a^*_{i_k}, b^*_{i_k}, f^*_{i_k}(t)) \in \mathcal{L}_{i_k}$,
\begin{align}
&a'_{i_k} \geq a^*_{i_k}\\
&f'_{i_k}(t) \leq \max_{L^*_{i_k} \in \mathcal{L}_{i_k}} f^*_{i_k}(t), \quad \forall t \in [a'_{i_k} + s_{i_k}, l_{i_k} + s_{i_k}].
\end{align}
Finally, the route with the minimum duration and the optimal charging decisions (i.e. when and how much to charge) is obtained. This process is repeated for a limited number of iterations ($OptIter$). This procedure corresponds to line \ref{alg.ts.7} - \ref{alg.ts.9} in Algorithm \ref{alg:tb1}.


\subsection{Perturbation}
If no improvement is achieved after a certain number of iterations ($ShakeTenure$), we randomly perform one of the relocate, cross or exchange moves between two different routes. After performing such random moves over a predefined number of iterations ($ShakeIter$), the tabu tenure ($TabuTenure$) is multiplied by a parameter $\tilde{\beta}$ which is less than 1 to enable the algorithm to explore a larger neighborhood. This procedure corresponds to line \ref{alg.ts.10} - \ref{alg.ts.17} in Algorithm \ref{alg:tb1}.

\begin{algorithm}[!h]
\caption{{\em Tabu Search()}} \label{alg:tb1}
\begin{small}
\begin{algorithmic}[1]
 \STATE Generate initial solution $s$; \label{alg.ts.1}
 \STATE Set best solution $s^* = s$;
 \STATE Initialize TabuList, $iter = 0, nonImp = 0, inonImp = 0$;
 \WHILE{$iter \leq MaxIter$ }
 \STATE Performing one of the three types of moves with the largest cost reduction which is not tabu in incumbent solution $s$; \label{alg.ts.5}
 \STATE Update the TabuList;
 \IF{$iter \% OptIter = 0$} \label{alg.ts.7}
 \STATE Optimizing charging decisions;
 \ENDIF \label{alg.ts.9}
 \IF{$nonImp \geq ShakeTenure$}\label{alg.ts.10}
 \STATE Performing a random move from the relocate, cross or exchange moves between two different routes; \label{alg.ts.11}
 \IF{$inonImp \geq ShakeIter$}
 \STATE $TabuTenure = \tilde{\beta} * TabuTenure$;
 \STATE $inonImp = 0$;
 \ENDIF
 \STATE $inonImp \leftarrow inonImp + 1$;
 \ENDIF\label{alg.ts.17}
 \IF{$cost(s) < cost(s^*)$}
 \STATE $s^* \leftarrow s$;
 \STATE $nonImp = 0$;
 \ELSE
 \STATE $nonImp \leftarrow nonImp + 1$;
 \ENDIF
 \STATE $iter \leftarrow iter + 1$;
 \ENDWHILE
 \STATE Return $s^*$;
\end{algorithmic}
\end{small}
\end{algorithm}

\section{Computational Experiments}
\label{sec:experiment}
In this section, we first test the performance of the proposed exact algorithm where the LP relaxation of the RMP is solved by ILOG CPLEX 12.63 with its default settings. Then, we evaluate the solution quality of the tabu search heuristic proposed in section \ref{sec:heuristic}. Both the exact algorithm and the tabu serach  are implemented in Java with Sun Oracles JDK 1.7.0 and the experiments are carried out on an Intel Xeon E5-1607 with a 3.10 GHz (Quad Core) CPU and 64G RAM running the Windows 10 operating system.

\subsection{Test instances}
We test our algorithms on the 120 instances generated by \citet{montoya2017electric}. The instances name is defined as {$tc\nu_1c\nu_2s\nu_3c\nu_4\nu_5$}. The parameter $\nu_1$ indicates the instance topology, i.e. random $\nu_1=0$, clustered $\nu_1 = 1$ and randomly clustered $\nu_1 = 2$). The parameter $\nu_2$ indicates the number of customers included in a instance, i.e. $\nu_2 = 10,20,40,80,160,320$. The parameter $\nu_3$ indicates the number of CSs. The parameter $\nu_4$ indicates the location of the CSs, i.e. it is set to $t$ when the CSs locations are determined using the p-median method, and set to $f$ when the locations are set randomly. The parameter $\nu_5$ gives the instance number, i.e. $\nu_5 = 0,1,2,3,4$. Furthermore, three types of piecewise linear functions (i.e. slow, moderate and fast) are employed to approximate the nonlinear charging functions. For more details on the instances, we refer to \citet{montoya2017electric}.

\subsection{Results of the exact algorithm}
We run the exact algorithm on instances with 10, 20 and 40 customers, and compare the results with state-of-the art results from the literature obtained by \citet{montoya2017electric} and \citet{froger2019improved}. In addition, we examine the impact of the subset-row inequalities when solving the E-VRP-NL. The tables show therefore results for the exact algorithm with SR cuts (BPC), and the exact algorithm without SR cuts (BP). We set the time limit to 3 hours for all instances.

The results of the comparison with \citet{montoya2017electric} are reported in tables \ref{tab:c10}-\ref{tab:c40}. The first column gives the name of the instance. Column ({\em Best}) presents the best solution found by either the exact algorithm BP or BPC. The next two groups of columns report the detailed results of both BP and BPC, respectively. They include the optimal cost of the set-partitioning model at the root node ({\em LP Cost}), the best upper bound found ({\em IP Cost}), the computational time at the root node ({\em Root Time}), the total computational time ({\em IP Time}), the number of SR cuts added ({\em SR Cuts}) and the number of nodes explored in the branch-and-bound tree ({\em Nodes}). The last two columns report the best solutions obtained by \citet{montoya2017electric} using a heuristic they named ILS+HC. Column ({\em BKS}) provides the best known solution found by the ILS+HC heuristic or by the Gurobi solver. Column ({\em Optimal}) denotes whether the solution is proven to be optimal. The instances solved to proven optimality by our algorithm are indicated in bold.

First we observe that adding SR cuts improves the performance of the exact algorithm. In fact,  5 more instances with 20 customers and 3 more instances with 40 customers are solved to proven optimality when including SR cuts. In addition, solution time is also improved on average when adding SR cuts. The average solution time when including SR cuts is 48.12s, 3832.08s and 8914.42s, against  53.40s, 5757.81s and 9965.18s for instances with 10, 20 and 40 customers, respectively, when SR cuts are excluded. Additionally, the number of nodes explored in the branching tree is also reduced when adding SR cuts from, on average, 285.80, 598.85 and 36.59 nodes to 2.20, 19.94 and 3.18 for the 10-, 20- and 40- customer instances, respectively. Secondly, our algorithm solves 20, 18 and 5 instances with 10, 20, 40 customers to optimality, respectively. Hence, outperforming state-of-the art solution methods available in the literature, i.e. the algorithm proposed by \citet{montoya2017electric} can only solve 7 instances with 20 customers to proven optimality, and none with 40 customers.

\begin{table}[htbp]
  \centering
  \caption{Computational Results of Instances with 10 Customers}
  \scalebox{0.6}{
    \begin{tabular}{ccccccccccccccccc}
    \toprule
    \multirow{2}[4]{*}{Instance} & \multirow{2}[4]{*}{Best} & \multicolumn{6}{c}{BPC}                       &       & \multicolumn{5}{c}{BP}                &       & \multicolumn{2}{c}{{\citet{montoya2017electric}}} \\
\cmidrule{3-8}\cmidrule{10-14}\cmidrule{16-17}          &       & LP Cost & IP Cost & Root Time (s) & IP Time (s) & SR Cuts & Nodes &       & LP Cost & IP Cost & Root Time (s) & IP Time (s) & Nodes &       & BKS   & Optimal \\
\cmidrule{1-8}\cmidrule{10-14}\cmidrule{16-17}    \textbf{tc2c10s2cf0} & 21.77  & 17.41  & 21.77  & 0.34  & 0.62  & 4     & 3     &       & 17.41  & 21.77  & 0.36  & 21.22  & 201   &       & 21.77 & Yes \\
    \textbf{tc0c10s2cf1} & 19.75  & 19.75  & 19.75  & 0.35  & 0.36  & 0     & 1     &       & 19.75  & 19.75  & 0.38  & 0.39  & 1     &       & 19.75 & Yes \\
    \textbf{tc1c10s2cf2} & 9.03  & 8.61  & 9.03  & 0.61  & 24.80  & 85    & 3     &       & 8.61  & 9.03  & 0.71  & 3.31  & 7     &       & 9.03  & Yes \\
    \textbf{tc1c10s2cf3} & 16.37  & 15.30  & 16.37  & 0.62  & 0.86  & 0     & 3     &       & 15.30  & 16.37  & 0.55  & 0.73  & 3     &       & 16.37 & Yes \\
    \textbf{tc1c10s2cf4} & 16.10  & 16.10  & 16.10  & 0.59  & 0.60  & 0     & 1     &       & 16.10  & 16.10  & 0.54  & 0.56  & 1     &       & 16.1  & Yes \\
    \textbf{tc2c10s2ct0} & 12.45  & 10.01  & 12.45  & 0.89  & 9.81  & 44    & 3     &       & 10.01  & 12.45  & 0.78  & 190.34  & 2237  &       & 12.45 & Yes \\
    \textbf{tc0c10s2ct1} & 12.30  & 12.27  & 12.30  & 0.52  & 0.84  & 0     & 3     &       & 12.27  & 12.30  & 0.53  & 0.84  & 3     &       & 12.3  & Yes \\
    \textbf{tc1c10s2ct2} & 10.75  & 9.89  & 10.75  & 1.07  & 60.56  & 53    & 3     &       & 9.89  & 10.75  & 0.98  & 23.56  & 29    &       & 10.75 & Yes \\
    \textbf{tc1c10s2ct3} & 13.17  & 13.17  & 13.17  & 0.44  & 0.45  & 0     & 1     &       & 13.17  & 13.17  & 0.42  & 0.44  & 1     &       & 13.17 & Yes \\
    \textbf{tc1c10s3ct4} & 13.21  & 13.21  & 13.21  & 0.65  & 0.66  & 0     & 1     &       & 13.21  & 13.21  & 0.62  & 0.64  & 1     &       & 13.21 & Yes \\
    \textbf{tc2c10s3cf0} & 21.77  & 17.41  & 21.77  & 0.38  & 0.72  & 4     & 3     &       & 17.41  & 21.77  & 0.38  & 22.00  & 203   &       & 21.77 & Yes \\
    \textbf{tc0c10s3cf1} & 19.75  & 19.75  & 19.75  & 0.37  & 0.38  & 0     & 1     &       & 19.75  & 19.75  & 0.33  & 0.34  & 1     &       & 19.75 & Yes \\
    \textbf{tc1c10s3cf2} & 9.03  & 8.61  & 9.03  & 0.66  & 20.84  & 76    & 3     &       & 8.61  & 9.03  & 0.58  & 3.04  & 7     &       & 9.03  & Yes \\
    \textbf{tc1c10s3cf3} & 16.37  & 15.30  & 16.37  & 0.57  & 1.01  & 17    & 3     &       & 15.30  & 16.37  & 0.57  & 5.30  & 27    &       & 16.37 & Yes \\
    \textbf{tc1c10s3cf4} & 14.90  & 14.76  & 14.90  & 0.56  & 0.74  & 0     & 3     &       & 14.76  & 14.90  & 0.54  & 0.71  & 3     &       & 14.9  & Yes \\
    \textbf{tc2c10s3ct0} & 11.51  & 9.22  & 11.51  & 1.35  & 241.84  & 50    & 3     &       & 9.22  & 11.51  & 1.30  & 785.38  & 2985  &       & 11.51 & Yes \\
    \textbf{tc0c10s3ct1} & 10.80  & 10.80  & 10.80  & 0.54  & 0.55  & 0     & 1     &       & 10.80  & 10.80  & 0.57  & 0.58  & 1     &       & 10.8  & Yes \\
    \textbf{tc1c10s3ct2} & 9.20  & 9.12  & 9.20  & 1.49  & 595.29  & 43    & 3     &       & 9.12  & 9.20  & 1.62  & 7.14  & 3     &       & 9.2   & Yes \\
    \textbf{tc1c10s3ct3} & 13.02  & 13.02  & 13.02  & 0.91  & 0.92  & 0     & 1     &       & 13.02  & 13.02  & 0.97  & 0.99  & 1     &       & 13.02 & Yes \\
    \textbf{tc1c10s2ct4} & 13.83  & 13.83  & 13.83  & 0.52  & 0.53  & 0     & 1     &       & 13.83  & 13.83  & 0.56  & 0.57  & 1     &       & 13.83 & Yes \\
    \hline
    Average & 14.25  & 13.38  & 14.25  & 0.67  & 48.12  & 18.80  & 2.20  &       & 13.38  & 14.25  & 0.67  & 53.40  & 285.80  &       & 14.25  &  \\
    \bottomrule
    \end{tabular}%
    }
  \label{tab:c10}%
\end{table}%

\begin{table}[htbp]
  \centering
  \caption{Computational Results of Instances with 20 Customers}
  \scalebox{0.6}{
    \begin{tabular}{ccccccccccccccccc}
    \toprule
    \multirow{2}[4]{*}{Instance} & \multirow{2}[4]{*}{Best} & \multicolumn{6}{c}{BPC}                       &       & \multicolumn{5}{c}{BP}                &       & \multicolumn{2}{c}{{\citet{montoya2017electric}}} \\
\cmidrule{3-8}\cmidrule{10-14}\cmidrule{16-17}             &       & LP Cost & IP Cost & Root Time (s) & IP Time (s) & SR Cuts & Nodes &       & LP Cost & IP Cost & Root Time (s) & IP Time (s) & Nodes &       & BKS   & Optimal \\
\cmidrule{1-8}\cmidrule{10-14}\cmidrule{16-17}    \textbf{tc2c20s3cf0} & 24.68  & 21.52  & 24.68  & 5.11  & 973.33  & 898   & 21    &       & 21.52  & 24.68  & 5.11  & 10800.00  & 4164  &       & 24.68  & No \\
    \textbf{tc1c20s3cf1} & 17.49  & 17.28  & 17.49  & 90.84  & 349.73  & 18    & 3     &       & 17.28  & 17.49  & 89.36  & 1628.25  & 21    &       & 17.50  & No \\
    \textbf{tc0c20s3cf2} & 27.47  & 26.34  & 27.47  & 3.54  & 804.45  & 2303  & 53    &       & 26.34  & 27.47  & 3.50  & 901.65  & 147   &       & 27.60  & No \\
    \textbf{tc1c20s3cf3} & 16.44  & 14.33  & 16.44  & 6.29  & 7660.03  & 89    & 3     &       & 14.33  & -     & 6.36  & 10800.00  & 631   &       & 16.63  & No \\
    \textbf{tc1c20s3cf4} & 17.00  & 17.00  & 17.00  & 2.44  & 2.51  & 0     & 1     &       & 17.00  & 17.00  & 2.37  & 2.38  & 1     &       & 17.00  & Yes \\
    \textbf{tc2c20s3ct0} & 25.79  & 24.21  & -     & 24.69  & 10800.00  & 1203  & 25    &       & 24.21  & 25.79  & 24.15  & 6240.64  & 375   &       & 25.79  & Yes \\
    tc1c20s3ct1 & 19.40  & -     & -     & -     & -     & -     & -     &       & 18.35  & 19.40  & 959.02  & 10800.00  & 25    &       & 18.95  & No \\
    \textbf{tc0c20s3ct2} & 17.08  & 15.70  & 17.08  & 5.18  & 1904.80  & 125   & 3     &       & 15.70  & 17.08  & 5.32  & 10800.00  & 801   &       & 17.08  & No \\
    \textbf{tc1c20s3ct3} & 12.60  & 12.35  & -     & 31.22  & 10800.00  & 11    & 2     &       & 12.35  & 12.60  & 31.23  & 1879.47  & 65    &       & 12.65  & No \\
    \textbf{tc1c20s3ct4} & 16.21  & 16.21  & 16.21  & 146.28  & 146.31  & 0     & 1     &       & 16.21  & 16.21  & 143.08  & 143.09  & 1     &       & 16.21  & Yes \\
    \textbf{tc2c20s4cf0} & 24.67  & 21.51  & 24.67  & 6.16  & 4266.91  & 5276  & 77    &       & 21.51  & 24.67  & 6.17  & 9120.37  & 3059  &       & 24.67  & No \\
    \textbf{tc1c20s4cf1} & 16.38  & 16.12  & -     & 46.33  & 10800.00  & 28    & 1     &       & 16.12  & 16.38  & 46.15  & 4272.03  & 51    &       & 16.39  & No \\
    \textbf{tc0c20s4cf2} & 27.47  & 26.28  & 27.47  & 6.54  & 1470.92  & 3259  & 77    &       & 26.28  & 27.47  & 6.37  & 2791.29  & 613   &       & 27.48  & No \\
    \textbf{tc1c20s4cf3} & 16.44  & 14.33  & 16.44  & 7.21  & 6239.09  & 94    & 3     &       & 14.33  & -     & 7.12  & 10800.00  & 530   &       & 16.56  & No \\
    \textbf{tc1c20s4cf4} & 17.00  & 17.00  & 17.00  & 4.91  & 5.21  & 0     & 1     &       & 17.00  & 17.00  & 4.76  & 4.77  & 1     &       & 17.00  & Yes \\
    \textbf{tc2c20s4ct0} & 26.02  & 25.19  & 26.02  & 27.50  & 10372.35  & 2948  & 81    &       & 25.19  & 26.02  & 27.43  & 1737.21  & 97    &       & 26.02  & No \\
    tc1c20s4ct1 & 18.26  & -     & -     & -     & -     & -     & -     &       & 17.22  & 18.26  & 510.68  & 10800.00  & 26    &       & 18.25  & Yes \\
    \textbf{tc0c20s4ct2} & 16.99  & 15.59  & 16.99  & 7.32  & 1541.75  & 135   & 3     &       & 15.59  & 17.13  & 7.18  & 10800.00  & 782   &       & 16.99  & No \\
    \textbf{tc1c20s4ct3} & 14.43  & 13.22  & 14.43  & 23.06  & 804.43  & 52    & 3     &       & 13.22  & 14.64  & 23.10  & 10800.00  & 586   &       & 14.43  & Yes \\
    \textbf{tc1c20s4ct4} & 17.00  & 17.00  & 17.00  & 35.61  & 35.62  & 0     & 1     &       & 17.00  & 17.00  & 35.04  & 35.06  & 1     &       & 17.00  & Yes \\
    \hline
    Average & 19.44  & 18.40  & 19.76  & 26.68  & 3832.08  & 913.28  & 19.94  &       & 18.34  & 19.79  & 97.17  & 5757.81  & 598.85  &       & 19.44  &  \\
    \bottomrule
    \end{tabular}%
    }
  \label{tab:c20}%
\end{table}%

\begin{table}[htbp]
  \centering
  \caption{Computational Results of Instances with 40 Customers}
  \scalebox{0.6}{
    \begin{tabular}{ccccccccccccccccc}
    \toprule
    \multirow{2}[4]{*}{Instance} & \multirow{2}[4]{*}{Best} & \multicolumn{6}{c}{BPC}                       &       & \multicolumn{5}{c}{BP}                &       & \multicolumn{2}{c}{{\citet{montoya2017electric}}} \\
\cmidrule{3-8}\cmidrule{10-14}\cmidrule{16-17}          &       & LP Cost & IP Cost & Root Time (s) & IP Time (s) & SR Cuts & Nodes &       & LP Cost & IP Cost & Root Time (s) & IP Time (s) & Nodes &       & BKS   & Optimal \\
\cmidrule{1-8}\cmidrule{10-14}\cmidrule{16-17}    \textbf{tc0c40s5cf0} & 32.20  & 31.00  & 32.20  & 16.24  & 2190.80  & 138   & 3     &       & 31.00  & 32.20  & 16.12  & 10800.00  & 156   &       & 32.67  & No \\
    tc1c40s5cf1 & 64.99  & 62.16  & 64.99  & 56.38  & 10800.00  & 1480  & 20    &       & 62.16  & -     & 55.29  & 10800.00  & 160   &       & 65.16  & No \\
    tc2c40s5cf2 & 27.95  & 26.19  & -     & 155.14  & 10800.00  & 58    & 1     &       & 26.19  & 27.95  & 151.28  & 10800.00  & 73    &       & 27.54  & No \\
    tc2c40s5cf3 & -     & 17.93  & -     & 1568.05  & 10800.00  & 53    & 2     &       & 17.93  & -     & 1529.81  & 10800.00  & 4     &       & 19.74  & No \\
    \textbf{tc0c40s5cf4} & 30.25  & 29.74  & 30.25  & 222.46  & 10713.94  & 53    & 3     &       & 29.74  & -     & 245.53  & 10800.00  & 23    &       & 30.77  & No \\
    \textbf{tc0c40s5ct0} & 27.91  & 27.73  & 27.91  & 70.06  & 482.29  & 10    & 3     &       & 27.73  & 27.91  & 68.37  & 1159.41  & 7     &       & 28.72  & No \\
    tc1c40s5ct1 & 52.42  & 51.10  & 52.42  & 113.26  & 10800.00  & 448   & 8     &       & 51.10  & -     & 109.74  & 10800.00  & 50    &       & 52.68  & No \\
    tc2c40s5ct2 & -     & 25.43  & -     & 258.91  & 10800.00  & 165   & 1     &       & 25.43  & -     & 248.82  & 10800.00  & 23    &       & 26.91  & No \\
    tc2c40s5ct3 & -     & 20.95  & -     & 6786.63  & 10800.00  & 0     & 1     &       & 20.95  & -     & 6385.76  & 10800.00  & 1     &       & 23.54  & No \\
    tc0c40s5ct4 & -     & -     & -     & -     & -     & -     & -     &       & -     & -     & -     & -     & -     &       & 28.63  & No \\
    \textbf{tc0c40s8cf0} & 30.40  & 28.44  & 30.40  & 71.22  & 7216.68  & 125   & 3     &       & 28.44  & -     & 69.54  & 10800.00  & 37    &       & 31.28  & No \\
    \textbf{tc1c40s8cf1} & 40.64  & 40.42  & 40.64  & 88.55  & 1341.39  & 32    & 3     &       & 40.42  & 40.64  & 87.31  & 6248.60  & 29    &       & 40.75  & No \\
    tc2c40s8cf2 & -     & 25.54  & -     & 138.22  & 10800.00  & 91    & 1     &       & 25.54  & -     & 136.04  & 10800.00  & 39    &       & 27.15  & No \\
    tc2c40s8cf3 & -     & 17.93  & -     & 1586.59  & 10800.00  & 55    & 2     &       & 17.93  & -     & 1536.20  & 10800.00  & 4     &       & 19.66  & No \\
    tc0c40s8cf4 & -     & 27.80  & -     & 250.49  & 10800.00  & 26    & 1     &       & 27.80  & -     & 245.14  & 10800.00  & 3     &       & 29.32  & No \\
    tc0c40s8ct0 & -     & 25.48  & -     & 123.43  & 10800.00  & 57    & 1     &       & 25.48  & -     & 118.61  & 10800.00  & 12    &       & 26.35  & No \\
    tc1c40s8ct1 & -     & -     & -     & -     & -     & -     & -     &       & -     & -     & -     & -     & -     &       & 40.56  & No \\
    tc2c40s8ct2 & -     & 24.45  & -     & 4931.59  & 10800.00  & 105   & 1     &       & 24.45  & -     & 4792.05  & 10800.00  & 1     &       & 26.33  & No \\
    tc2c40s8ct3 & -     & 20.32  & -     & 10800.00  & 10800.00  & 0     & 0     &       & 20.32  & -     & 10800.00  & 10800.00  & 0     &       & 22.71  & No \\
    tc0c40s8ct4 & -     & -     & -     & -     & -     & -     & -     &       & -     & -     & -     & -     & -     &       & 29.20  & No \\
    \hline
    Average & 38.34  & 29.57  & 39.83  & 1602.19  & 8914.42  & 170.35  & 3.18  &       & 29.57  & 32.17  & 1564.45  & 9965.18  & 36.59  &       & 31.48  &  \\
    \bottomrule
    \end{tabular}%
    }
  \label{tab:c40}%
\end{table}%
The results of the comparison with \citet{froger2019improved} are reported in tables \ref{tab:exactcompare10} and \ref{tab:exactcompare20} for instances with 10 and 20 customers. Three models, including Node-, Arc- and Path-based models, are proposed by \citet{froger2019improved}. According to their results, the path-based model outperforms both the Node- and Arc-based models. Therefore, we only compare our results with those obtained by solving the Path-based model. In tables \ref{tab:exactcompare10} and \ref{tab:exactcompare20}, three groups of columns are shown for the BP, BPC algorithm and the Path-model of \citet{froger2019improved}. For each group of columns, we report the best upper bound ({\em Obj}), the lower bound at the root node ({\em LP}) and the computational time in seconds ({\em Time}). The last column {\em Optimal} indicates  whether an instance is optimally solved by the path-model.
We observe that instances with 10 customers are solved to optimality by both the BP and BPC, and the path-mode of \citet{froger2019improved}. For instances with 20 customers, our exact algorithm (BP and BPC) clearly outperforms the path-model of \citet{froger2019improved} that is able to solve only 5 instances optimally. In addition, BPC and BP provide much better lower bounds at the root node and have much shorter computation times.

\begin{table}[htbp]
  \centering
   \caption{Comparison of results of the Exact Algorithms and Path-based model in \citet{froger2019improved} (10-customers Instances)}
  \scalebox{0.7}{
    \begin{tabular}{ccccccccccccr}
    \toprule
    \multirow{2}[4]{*}{Instance} & \multicolumn{3}{c}{BPC} &       & \multicolumn{3}{c}{BP} &       & \multicolumn{4}{c}{Path Model} \\
\cmidrule{2-4}\cmidrule{6-8}\cmidrule{10-13}          & Obj   & LP    &  Time &       & Obj   & LP    & Time  &       & Obj   & LP    & Time  & Optimal\\
    \midrule
    tc2c10s2cf0 & 21.77  & 17.41  & 0.62  &       & 21.77  & 17.41  & 21.22  &       & 21.77  & 10.96  & 348.00  & Yes \\
    tc0c10s2cf1 & 19.75  & 19.75  & 0.36  &       & 19.75  & 19.75  & 0.39  &       & 19.75  & 12.26  & 11.00  &Yes \\
    tc1c10s2cf2 & 9.03  & 8.61  & 24.80  &       & 9.03  & 8.61  & 3.31  &       & 9.03  & 6.81  & 4.00  & Yes \\
    tc1c10s2cf3 & 16.37  & 15.30  & 0.86  &       & 16.37  & 15.30  & 0.73  &       & 16.37  & 11.26  & 56.00  & Yes\\
    tc1c10s2cf4 & 16.10  & 16.10  & 0.60  &       & 16.10  & 16.10  & 0.56  &       & 16.10  & 12.74  & 10.00  & Yes\\
    tc2c10s2ct0 & 12.45  & 10.01  & 9.81  &       & 12.45  & 10.01  & 190.34  &       & 12.45  & 4.94  & 794.00  & Yes\\
    tc0c10s2ct1 & 12.30  & 12.27  & 0.84  &       & 12.30  & 12.27  & 0.84  &       & 12.30  & 9.36  & 10.00  & Yes\\
    tc1c10s2ct2 & 10.75  & 9.89  & 60.56  &       & 10.75  & 9.89  & 23.56  &       & 10.75  & 6.99  & 118.00  & Yes\\
    tc1c10s2ct3 & 13.17  & 13.17  & 0.45  &       & 13.17  & 13.17  & 0.44  &       & 13.17  & 9.01  & 4.00  & Yes\\
    tc1c10s3ct4 & 13.21  & 13.21  & 0.66  &       & 13.21  & 13.21  & 0.64  &       & 13.21  & 8.99  & 10.00  & Yes\\
    tc2c10s3cf0 & 21.77  & 17.41  & 0.72  &       & 21.77  & 17.41  & 22.00  &       & 21.77  & 10.96  & 350.00  & Yes\\
    tc0c10s3cf1 & 19.75  & 19.75  & 0.38  &       & 19.75  & 19.75  & 0.34  &       & 19.75  & 12.26  & 11.00  & Yes\\
    tc1c10s3cf2 & 9.03  & 8.61  & 20.84  &       & 9.03  & 8.61  & 3.04  &       & 9.03  & 6.81  & 4.00  & Yes \\
    tc1c10s3cf3 & 16.37  & 15.30  & 1.01  &       & 16.37  & 15.30  & 5.30  &       & 16.37  & 11.25  & 48.00  & Yes \\
    tc1c10s3cf4 & 14.90  & 14.76  & 0.74  &       & 14.90  & 14.76  & 0.71  &       & 14.90  & 11.17  & 13.00  & Yes \\
    tc2c10s3ct0 & 11.51  & 9.22  & 241.84  &       & 11.51  & 9.22  & 785.38  &       & 11.51  & 4.41  & 2716.00  & Yes\\
    tc0c10s3ct1 & 10.80  & 10.80  & 0.55  &       & 10.80  & 10.80  & 0.58  &       & 10.80  & 8.25  & 7.00  & Yes \\
    tc1c10s3ct2 & 9.20  & 9.12  & 595.29  &       & 9.20  & 9.12  & 7.14  &       & 9.20  & 6.61  & 45.00  & Yes \\
    tc1c10s3ct3 & 13.02  & 13.02  & 0.92  &       & 13.02  & 13.02  & 0.99  &       & 13.02  & 7.33  & 27.00  & Yes \\
    tc1c10s2ct4 & 13.83  & 13.83  & 0.53  &       & 13.83  & 13.83  & 0.57  &       & 13.83  & 11.15  & 4.00  & Yes \\
    \hline
    Average   & 14.25  & 13.38  & 48.12  &       & 14.25  & 13.38  & 53.40  &       & 14.25  & 9.18  & 229.50  &  \\
    \bottomrule
    \end{tabular}}
  \label{tab:exactcompare10}%
\end{table}%

\begin{table}[htbp]
  \centering
  \caption{Comparison of results of the Exact Algorithms and Path-based model in \citet{froger2019improved}
  (20-customers Instances)}
  \scalebox{0.7}{
    \begin{tabular}{ccccccccccccc}
        \toprule
    \multirow{2}[4]{*}{Instance} & \multicolumn{3}{c}{BPC} &       & \multicolumn{3}{c}{BP} &       & \multicolumn{4}{c}{Path Model} \\
\cmidrule{2-4}\cmidrule{6-8}\cmidrule{10-13}          & Obj   & LP    &  Time &       & Obj   & LP    & Time  &       & Obj   & LP    & Time  & Optimal \\
    \midrule
    tc2c20s3cf0 & 24.68  & 21.52  & 973.33  &       & 24.68  & 21.52  & 10800.00  &       & 24.68  & 11.95  & 10800.00  & No \\
    tc1c20s3cf1 & 17.49  & 17.28  & 349.73  &       & 17.49  & 17.28  & 1628.25  &       & 17.49  & 14.08  & 10800.00  & No \\
    tc0c20s3cf2 & 27.47  & 26.34  & 804.45  &       & 27.47  & 26.34  & 901.65  &       & 27.49  & 15.34  & 10800.00  & No \\
    tc1c20s3cf3 & 16.44  & 14.33  & 7660.03  &       & -     & 14.33  & 10800.00  &       & 16.48  & 9.92  & 10800.00  & No \\
    tc1c20s3cf4 & 17.00  & 17.00  & 2.51  &       & 17.00  & 17.00  & 2.38  &       & 17.00  & 14.95  & 111.00  & Yes \\
    tc2c20s3ct0 & -     & 24.21  & 10800.00  &       & 25.79  & 24.21  & 6240.64  &       & 25.80  & 11.74  & 10800.00  & No \\
    tc1c20s3ct1 & -     & -     & -     &       & 19.40  & 18.35  & 10800.00  &       & 18.94  & 14.98  & 10800.00  & No \\
    tc0c20s3ct2 & 17.08  & 15.70  & 1904.80  &       & 17.08  & 15.70  & 10800.00  &       & 17.08  & 11.66  & 10800.00  & No \\
    tc1c20s3ct3 & -     & 12.35  & 10800.00  &       & 12.60  & 12.35  & 1879.47  &       & 12.60  & 9.39  & 1529.00  & Yes \\
    tc1c20s3ct4 & 16.21  & 16.21  & 146.31  &       & 16.21  & 16.21  & 143.09  &       & 16.21  & 13.91  & 163.00  & Yes \\
    tc2c20s4cf0 & 24.67  & 21.51  & 4266.91  &       & 24.67  & 21.51  & 9120.37  &       & 24.67  & 11.81  & 10800.00  & No \\
    tc1c20s4cf1 & -     & 16.12  & 10800.00  &       & 16.38  & 16.12  & 4272.03  &       & 16.38  & 12.58  & 10800.00  & No \\
    tc0c20s4cf2 & 27.47  & 26.28  & 1470.92  &       & 27.47  & 26.28  & 2791.29  &       & 27.47  & 15.34  & 10800.00  & No \\
    tc1c20s4cf3 & 16.44  & 14.33  & 6239.09  &       & -     & 14.33  & 10800.00  &       & 16.84  & 9.92  & 10800.00  & No \\
    tc1c20s4cf4 & 17.00  & 17.00  & 5.21  &       & 17.00  & 17.00  & 4.77  &       & 17.00  & 14.69  & 134.00  & Yes \\
    tc2c20s4ct0 & 26.02  & 25.19  & 10372.35  &       & 26.02  & 25.19  & 1737.21  &       & 26.02  & 11.46  & 10800.00  & No \\
    tc1c20s4ct1 & -     & -     & -     &       & 18.26  & 17.22  & 10800.00  &       & 18.02  & 13.72  & 10800.00  & No \\
    tc0c20s4ct2 & 16.99  & 15.59  & 1541.75  &       & 17.13  & 15.59  & 10800.00  &       & 16.99  & 11.40  & 10800.00  & No \\
    tc1c20s4ct3 & 14.43  & 13.22  & 804.43  &       & 14.64  & 13.22  & 10800.00  &       & 14.43  & 9.69  & 10800.00  & No \\
    tc1c20s4ct4 & 17.00  & 17.00  & 35.62  &       & 17.00  & 17.00  & 35.06  &       & 17.00  & 14.54  & 506.00  & Yes \\
    \hline
    Average   & 19.76  & 18.40  & 3832.08  &       & 19.79  & 18.34  & 5757.81  &       & 19.43  & 12.65  & 8222.15  &  \\
    \bottomrule
    \end{tabular}}
  \label{tab:exactcompare20}%
\end{table}%

\subsection{Heuristic results}
In this section, we assess the performance of the tabu search heuristic on the 120 instances of \citet{montoya2017electric}. In our experiments, the parameters $MaxIter$ and $TabuTenure$ are set depending on the size of the instance. The parameter $TabuTenure$ is determined by dividing the instance size by a parameter $\tilde{\alpha}$ that is set to be 2 or 3 based on preliminary experiments. Moreover, the parameters $ShakeIter$, $OptIter$ and $\tilde{\beta}$ are set to be equal to 20, 5 and 0.9, respectively, and the parameter $ShakeTenure$ is equal to 40, 80 or 160. To calibrate the value of $\tilde{\alpha}$ and $ShakeTenure$, we conduct $2\times3$ experiments in which all parameters are fixed except one. The gap between the objective cost and the baseline which is regarded as the minimum objective value with different parameter combinations for each instance is reported in Table \ref{tab:para}. Based on the results, the parameters settings are summarized in Table \ref{tab:parameter}.

\begin{table}[htbp]
  \centering
  \caption{Gap (\%) between baseline and objective cost of each combination of $\tilde{\alpha}$ and $ShakeTenure$}
  \scalebox{0.9}{
  \setlength{\tabcolsep}{6mm}
    \begin{tabular}{ccccc}
    \toprule
    \multirow{2}[4]{*}{Size} & \multirow{2}[4]{*}{$\alpha$} & \multicolumn{3}{c}{$ShakeTenure$} \\
\cmidrule{3-5}          &       & 40    & 80    & 160 \\
    \midrule
    \multirow{2}[1]{*}{10, 20} & 2     & 0.40  & \textbf{0.00 } & 0.10  \\
          & 3     & 0.33  & 0.10  & 0.06  \\
    \multirow{2}[0]{*}{40, 80} & 2     & 0.71  & 0.47  & 0.40  \\
          & 3     & 0.51  & 0.14  & \textbf{0.00 } \\
    \multirow{2}[1]{*}{160, 320} & 2     & 0.14  & 0.42  & 0.03  \\
          & 3     & 0.45  & 0.16  & \textbf{0.00 } \\
    \bottomrule
    \end{tabular}}
  \label{tab:para}%
\end{table}%

\begin{table}[htbp]
  \centering
  \caption{Parameters Settings}
  \scalebox{0.9}{
    \begin{tabular}{ccccccc}
    \toprule
    Size  & MaxIter & $\alpha$ & ShakeIter & ShakeTenure & OptIter & $\beta$ \\
    \midrule
    10,20 & 1000  & 2     &20& 80    & 5 & 0.9\\
    40,80 & 2000  & 3     &   20  & 160    &    5   & 0.9 \\
    160,320 & 4000  & 3     &  20    & 160   &   5    &  0.9\\
    \bottomrule
    \end{tabular}%
    }
  \label{tab:parameter}%
\end{table}%

We first examine the solution quality of the tabu search algorithm on small-scale instances. Table \ref{tab:comparison} compares the detailed results of the heuristic algorithm with the best solutions obtained by the BPC and BP algorithms on 10- and 20-customer instances. Column {\em Instance} gives the name of instances. The table reports the best costs found by the tabu search ({\em TS}) and by the exact algorithm ({\em Exact}) as well as the computational time in seconds for the tabu search ({\em T$_1$}(s)) and the exact algorithm ({\em T$_2$}(s)). The instances which are optimally solved by the tabu search algorithm are marked in bold. As shown in table \ref{tab:comparison}, the tabu search algorithm is able to solve all instances with 10 customers and 16 instances with 20 customers to optimality. In addition, the tabu search algorithm achieves better solutions for instances {\em tc1c20s3ct1} and {\em tc1c20s4ct1} which are not solved to optimality by the exact algorithm. Furthermore, the computational time required by the tabu search algorithm is much shorter, i.e. only 0.46s and 1.15s on average for 10- and 20-customer instances, respectively.
\begin{table}[htbp]
  \centering
  \caption{Comparison of the Results of the Exact and the Tabu Search Algorithm for 10- and 20-customer instances}
   \scalebox{0.7}{
    \begin{tabular}{cccccccccc}
    \toprule
    Instance & TS    & T$_1$(s) & Exact & T$_2$(s) & Instance & TS    & T$_1$(s) & Exact & T$_2$(s) \\
    \midrule
    \textbf{tc2c10s2cf0} & 21.77  & 0.73  & 21.77  & 0.62  & \textbf{tc2c20s3cf0} & 24.68  & 0.79  & 24.68  & 973.33  \\
    \textbf{tc0c10s2cf1} & 19.75  & 0.46  & 19.75  & 0.36  & \textbf{tc1c20s3cf1} & 17.49  & 1.57  & 17.49  & 349.73  \\
    \textbf{tc1c10s2cf2} & 9.03  & 0.51  & 9.03  & 3.31  & \textbf{tc0c20s3cf2} & 27.47  & 0.85  & 27.47  & 804.45  \\
    \textbf{tc1c10s2cf3} & 16.37  & 0.52  & 16.37  & 0.73  & tc1c20s3cf3 & 16.55  & 1.28  & 16.44  & 7660.03  \\
    \textbf{tc1c10s2cf4} & 16.10  & 0.41  & 16.10  & 0.56  & \textbf{tc1c20s3cf4} & 17.00  & 1.05  & 17.00  & 2.38  \\
    \textbf{tc2c10s2ct0} & 12.45  & 0.64  & 12.45  & 9.81  & \textbf{tc2c20s3ct0} & 25.79  & 0.97  & 25.79  & 6240.64  \\
    \textbf{tc0c10s2ct1} & 12.30  & 0.54  & 12.30  & 0.84  & tc1c20s3ct1 & 18.94  & 1.64  & 19.40  & - \\
    \textbf{tc1c10s2ct2} & 10.75  & 0.49  & 10.75  & 23.56  & \textbf{tc0c20s3ct2} & 17.08  & 0.95  & 17.08  & 1904.80  \\
    \textbf{tc1c10s2ct3} & 13.17  & 0.41  & 13.17  & 0.44  & \textbf{tc1c20s3ct3} & 12.60  & 1.16  & 12.60  & 1879.47  \\
    \textbf{tc1c10s3ct4} & 13.21  & 0.42  & 13.21  & 0.64  & \textbf{tc1c20s3ct4} & 16.21  & 1.21  & 16.21  & 143.09  \\
    \textbf{tc2c10s3cf0} & 21.77  & 0.25  & 21.77  & 0.72  & \textbf{tc2c20s4cf0} & 24.67  & 0.82  & 24.67  & 4266.91  \\
    \textbf{tc0c10s3cf1} & 19.75  & 0.31  & 19.75  & 0.34  & \textbf{tc1c20s4cf1} & 16.38  & 1.36  & 16.38  & 4272.03  \\
    \textbf{tc1c10s3cf2} & 9.03  & 0.41  & 9.03  & 3.04  & \textbf{tc0c20s4cf2} & 27.47  & 0.87  & 27.47  & 1470.92  \\
    \textbf{tc1c10s3cf3} & 16.37  & 0.42  & 16.37  & 1.01  & tc1c20s4cf3 & 16.62  & 1.24  & 16.44  & 6239.09  \\
    \textbf{tc1c10s3cf4} & 14.90  & 0.40  & 14.90  & 0.71  & \textbf{tc1c20s4cf4} & 17.00  & 1.11  & 17.00  & 4.77  \\
    \textbf{tc2c10s3ct0} & 11.51  & 0.43  & 11.51  & 241.84  & \textbf{tc2c20s4ct0} & 26.02  & 1.11  & 26.02  & 1737.21  \\
    \textbf{tc0c10s3ct1} & 10.80  & 0.40  & 10.80  & 0.55  & tc1c20s4ct1 & 18.04  & 1.96  & 18.26  & - \\
    \textbf{tc1c10s3ct2} & 9.20  & 0.56  & 9.20  & 7.14  & \textbf{tc0c20s4ct2} & 16.99  & 0.86  & 16.99  & 1541.75  \\
    \textbf{tc1c10s3ct3} & 13.02  & 0.47  & 13.02  & 0.92  & \textbf{tc1c20s4ct3} & 14.43  & 1.11  & 14.43  & 804.43  \\
    \textbf{tc1c10s2ct4} & 13.83  & 0.36  & 13.83  & 0.53  & \textbf{tc1c20s4ct4} & 17.00  & 1.18  & 17.00  & 35.06  \\
     \hline
    Average & 14.25  & 0.46  & 14.25  & 14.88  &       & 19.42  & 1.15  & 19.44  & 2240.56  \\
    \bottomrule
    \end{tabular}%
    }
  \label{tab:comparison}%
\end{table}%

According to \citet{montoya2017electric}, the routing and the charging decisions can be integrated and made either simultaneously, or consecutively when solving the E-VRP-NL. In the tabu search algorithm, these decision are integrated due to the use of the recursive functions introduced in section \ref{subsec:recursive}. This in opposite to existing approaches in the literature where a route first- charge second approach is applied. To evaluate the benefit of integrating routing and charging decisions, we report in table \ref{tab:heu} the improvements achieved by decision integration with regard to solution cost ($\Delta_{Cost}$), solving time ($\Delta_{Time}$) and the number of solved  instances ($\#I$) compared to \citet{montoya2017electric}. On average, the cost is reduced by $1.75\%$, and in 88 out of 120 instances the cost is improved. We note that the larger the instances the more significant are the improvements. For instances with 40 customers, 19 instances are improved. Moreover, all instances with 80,160 and 320 customers are improved. Next to solution cost, solution time is also reduced by up to $90.71\%$. Overall, the results indicate that integrating the routing and charging decisions improves both solution time and quality when solving the E-VRP-NL.
\begin{table}[htbp]
  \centering
  \caption{Results of \citet{montoya2017electric} improved by the Tabu Search Algorithms}
  \scalebox{0.9}{
    \begin{tabular}{cccccccc}
    \toprule
    Size  & 10    & 20    & 40    & 80    & 160   & 320   & \multicolumn{1}{c}{Avg} \\
    \midrule
    $\Delta_{Cost}$ & 0.00\% & -0.12\% & -0.23\% & -1.45\% & -2.91\% & -5.81\% & \multicolumn{1}{c}{-1.75\%} \\
    $\Delta_{Time}$ & -90.71\% & -87.05\% & -84.06\% & -74.19\% & -73.89\% & -81.92\% & \multicolumn{1}{c}{-81.97\%} \\
    \#I   & 0     & 9     & 19    & 20    & 20    & 20    &  \\
    \bottomrule
    \end{tabular}%
    }
  \label{tab:heu}%
\end{table}%

In \citet{froger2019improved}, 23 new best solutions are found by applying the labeling algorithm on each of the routes in the solutions reported in \citet{montoya2017electric}. We compare our results with these 23 solutions in table \ref{tab:compare2019}, where column {\em Instance} indicates the instances' name, column {\em Old BKS} presents the results in \citet{froger2019improved}. The produced results by the tabu search algorithm are reported in column {\em Our Results}, and compared with the results in \citet{froger2019improved}. Column  {\em $\Delta_c$} reports the gains achieved by the tabu search. Instances with an improved cost are marked in bold. The tabu search algorithm produces new best solutions for 21 out of 23 instances. The gains are more significant for larger instances.
\begin{table}[!h]
  \centering
  \caption{Comparison between the new best results reported in \citet{froger2019improved}}
  \scalebox{0.8}{
    \begin{tabular}{lccc}
    \toprule
    Instance & Old BKS & Our Results & $\Delta_c$ \\
    \midrule
    tc0c40s8cf0 & 31.045 & 31.172  & 0.411\% \\
    \textbf{tc2c40s8cf2} & 27.141 & 27.139  & -0.006\% \\
    tc2c40s5cf2 & 27.536 & 27.588  & 0.188\% \\
    \textbf{tc0c80s8cf1} & 45.225 & 44.972  & -0.559\% \\
    \textbf{tc1c80s12cf2} & 29.532 & 28.930  & -2.038\% \\
    \textbf{tc2c80s8cf4} & 49.171 & 48.253  & -1.868\% \\
    \textbf{tc2c80s8ct3} & 32.312 & 31.685  & -1.941\% \\
    \textbf{tc0c160s16cf4} & 82.863 & 79.576  & -3.966\% \\
    \textbf{tc0c160s16ct4} & 82.323 & 78.435  & -4.723\% \\
    \textbf{tc0c160s24ct4} & 80.796 & 78.915  & -2.328\% \\
    \textbf{tc0c160s24cf4} & 81.38 & 78.951  & -2.985\% \\
    \textbf{tc1c160s16cf3} & 71.509 & 69.162  & -3.282\% \\
    \textbf{tc1c160s24cf3} & 68.512 & 64.894  & -5.281\% \\
    \textbf{tc1c320s24cf2} & 152.063 & 142.365  & -6.377\% \\
    \textbf{tc1c320s24cf3} & 117.462 & 110.578  & -5.861\% \\
    \textbf{tc1c320s38cf2} & 141.62 & 135.631  & -4.229\% \\
    \textbf{tc1c320s38ct3} & 116.065 & 110.203  & -5.051\% \\
    \textbf{tc2c320s24cf0} & 182.453 & 166.110  & -8.957\% \\
    \textbf{tc2c320s24ct4} & 121.82 & 116.220  & -4.597\% \\
    \textbf{tc2c320s38cf4} & 122.318 & 114.631  & -6.285\% \\
    \textbf{tc2c320s38ct0} & 190.963 & 167.084  & -12.505\% \\
    \textbf{tc2c320s38ct1} & 94.533 & 88.901  & -5.958\% \\
    \textbf{tc2c320s38ct4} & 121.657 & 115.086  & -5.401\% \\
    \bottomrule
    \end{tabular}%
    }
  \label{tab:compare2019}%
\end{table}%

\subsection{The benefit of nonlinear charging functions}
\begin{figure}
\centering
\subfigure[Underestimated linear approximation]{
\label{fig.under}
\begin{minipage}[b]{0.4\textwidth}
\includegraphics[width=1\textwidth]{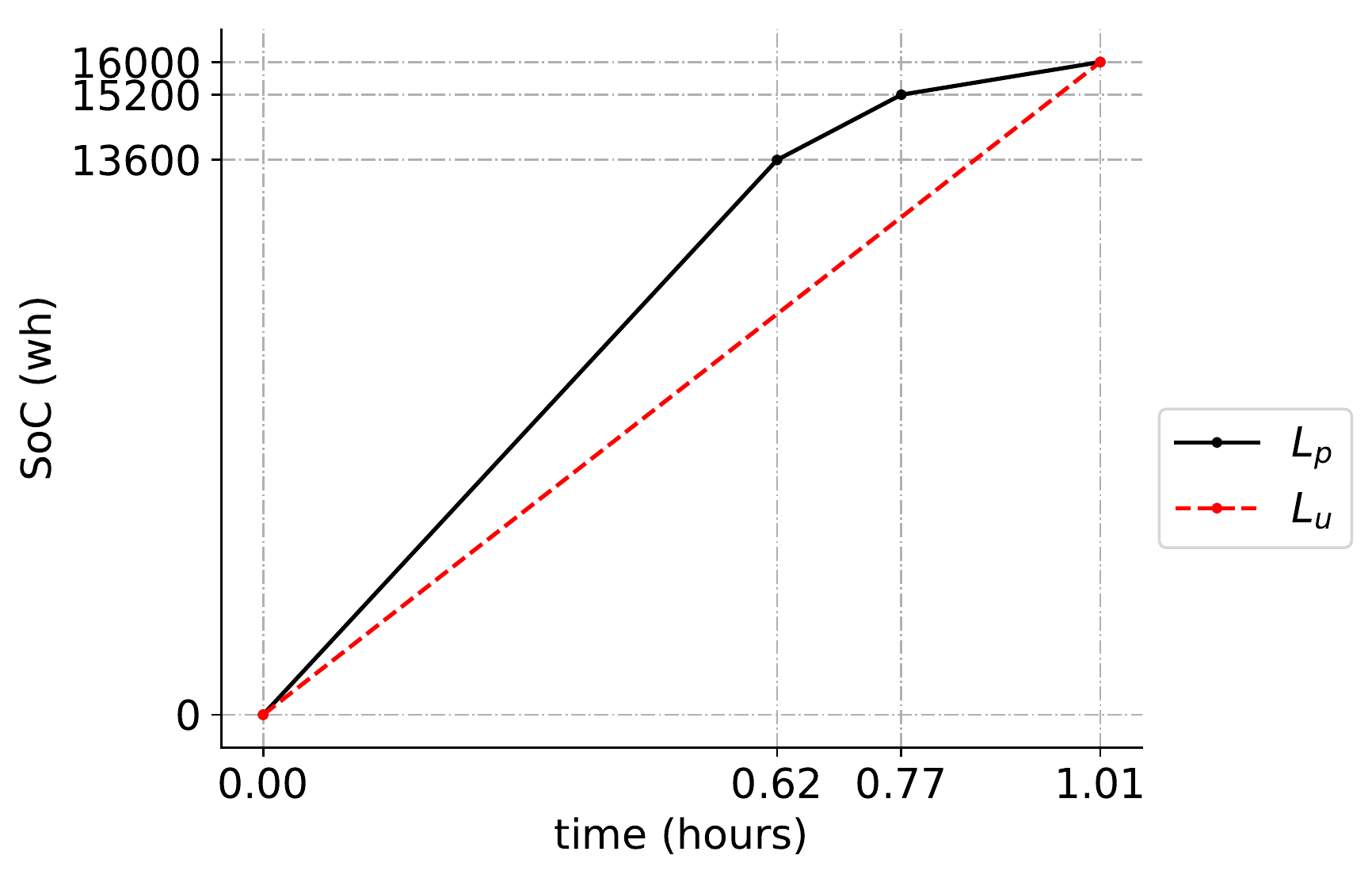}\\
\end{minipage}}
\subfigure[Overestimated linear approximation]{
\label{fig.over1}
\begin{minipage}[b]{0.4\textwidth}
\includegraphics[width=1\textwidth]{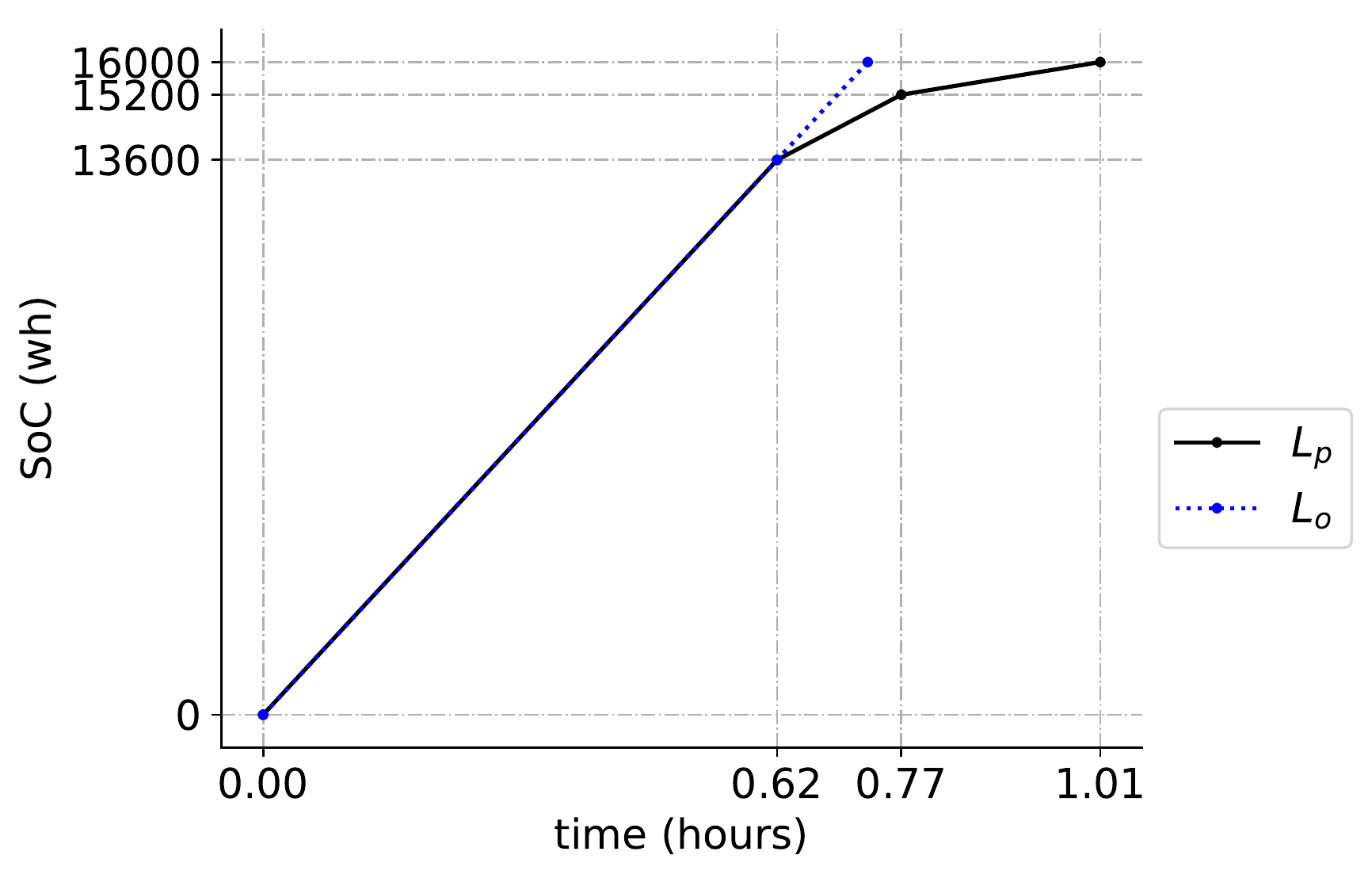}\\
\end{minipage}}
\caption{Piecewise linear function VS linear charging function}
\label{fig.linear}
\end{figure}

In this section, we investigate the benefit of adopting nonlinear charging functions approximated with piecewise linear functions in the electric vehicle routing problem. We compare the results obtained by using the piecewise linear functions with the case of linearly increasing charging functions. For this matter we use two different linear functions: First, a linear function $L_u$ based on the concave charging stage of the SoC that underestimates the real charging function (Figure \ref{fig.under}). Secondly, a linear function $L_o$ based on the linear first charging stage of the SoC that overestimates the real charging function (Figure \ref{fig.over1}). We run the BPC algorithm on instances with 10 customers where the charging functions are linear and evaluate the generated routes using the piecewise linear functions.

\begin{figure}
\centering
\includegraphics[scale = 0.5]{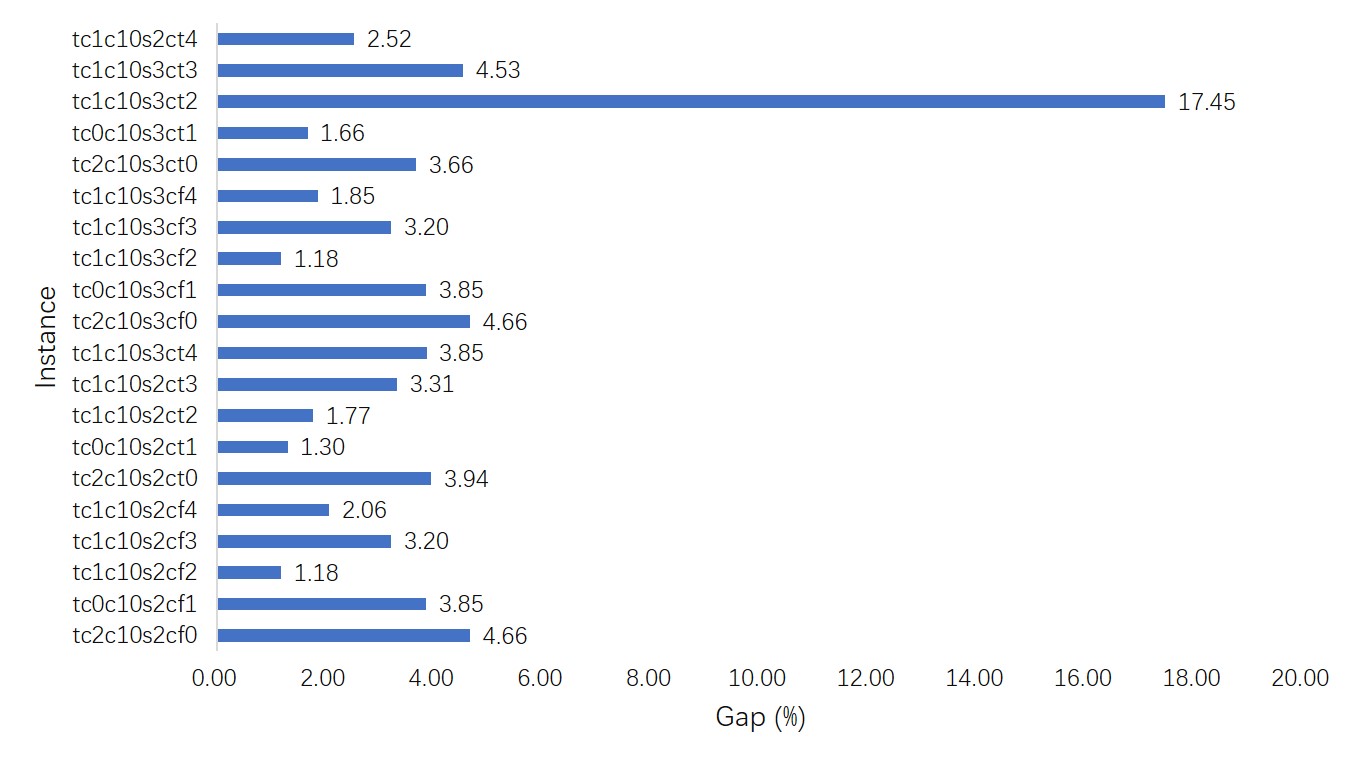}
\caption{Comparison on cost of instances with 10 customers using piecewise linear function and underestimate linear function}
\label{fig.comu}
\end{figure}

\begin{figure}
\centering
\includegraphics[scale = 0.45]{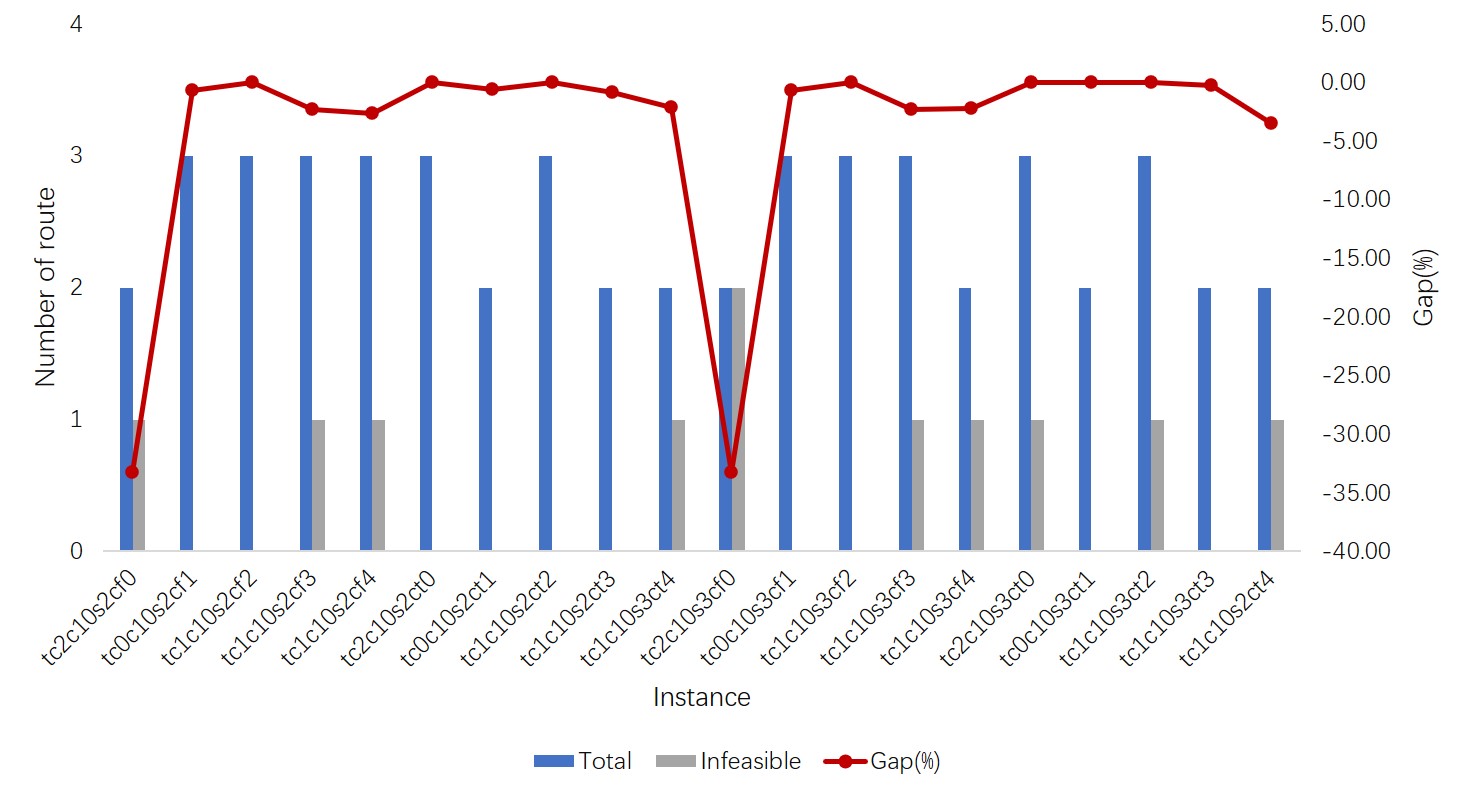}
\caption{Comparison on results of instances with 10 customers using piecewise linear function and overestimate linear function}
\label{fig.over}
\end{figure}

On the one hand, using an underestimating linear function increases the cost for all instances as shown by the results reported in Figure \ref{fig.comu}. This is because the underestimating linear charging functions lead to longer travel times because of the longer charging times. On the other hand, when using the overestimating linear charging function, charging times are underestimated and routes may be infeasible when evaluated using the piecewise linear charging functions as both battery levels and route durations may be violated. In Figure \ref{fig.over}, we present the total number of routes in the solutions ({\em Total}) against the number of routes that runed out to be infeasible ({\em Infeasible}). Additionally, the gaps ({\em Gap(\%))} between the cost of instances employing piecewise linear charging functions and the overestimating linear charging functions are reported in Figure \ref{fig.over}. Using overestimating linear charging functions results in lower costs.

\section{Conclusion}
EVs are a potential instrument to face environmental challenges. Yet, they suffer from technical weaknesses that prevent their wider uptake and render the management and planning of transportation operations a challenging task. This paper outlines exact and heuristic algorithms to solve the E-VRP-NL introduced by \citet{montoya2017electric}. The nonlinear battery charging process is hard to capture in optimization models, and hence is approximated by a linear piecewise function to render models tractable and facilitate the development of efficient algorithms. However, including the piecewise linear function significantly complicates the pricing problem where both routing and charging decisions must be handled simultaneously. We formulate the E-VRP-NL as a set-partitioning model. We introduce tailored recursive functions that capture both decisions and show how we can efficiently embed them in the underlying column generation algorithm. This results in an efficient exact algorithm for the E-VRP-NL. Next to the exact algorithm, a tabu search based heuristic is also developed to solve large instances quickly.

To assess the performance of the proposed algorithms, we test them on benchmark instances provided by \citet{montoya2017electric} and compare with state-of-the-art algorithms from the literature. The computational results demonstrate that our exact algorithm is capable of solving medium-scale instances to optimality, clearly outperforming existing tools. Furthermore, the proposed tabu search heuristic proves to be superior to existing heuristic algorithms from the literature. In fact, the tabu search heuristic obtain optimal solutions on small-scale instances and substantially improve the results from the literature on large-scale instances.

Because existing algorithms for the E-VRP-NL can only solve medium-size problems, there is room for improvement and multiple directions for future research are identified. First, current algorithms could be further improved by e.g., exploring the exploitation of problem specific valid inequalities and other techniques to accelerate the solution time. Second, other variants of the E-VRP-NL could be explored where e.g., the availability of CSs is probabilistic.
\label{sec:conclusion}

\section*{Acknowledgments}
This research was partially supported by the Young Elite Scientists Sponsorship Program by China Association for Science and Technology (Grants No. 2019QNRC001).

\bibliographystyle{ormsv080}
\bibliography{bpp}
\end{document}